\documentclass[prd,showpacs,superscriptaddress,altaffilletter,nofootinbib]{revtex4}
\usepackage[dvips]{graphicx}
\usepackage{amsmath,latexsym}

\newcommand{\be}{\begin{equation}}
\newcommand{\ee}{\end{equation}}
\newcommand{\bea}{\begin{eqnarray}}
\newcommand{\eea}{\end{eqnarray}}
\newcommand{\der}{\partial}
\newcommand{\vphi}{\varphi}

\begin{document}

\title{Brans-Dicke Galileon and the Variational Principle}

\author{Israel Quiros}\email{iquiros@fisica.ugto.mx}\affiliation{Dpto. Ingenier\'ia Civil, Divisi\'on de Ingenier\'ia, Universidad de Guanajuato, Gto., M\'exico.}

\author{Ricardo Garc\'{\i}a-Salcedo}\email{rigarcias@ipn.mx}\affiliation{CICATA - Legaria del Instituto Polit\'ecnico Nacional, 11500, M\'exico, D.F., M\'exico.}

\author{Tame Gonzalez}\email{tamegc72@gmail.com}\affiliation{Dpto. Ingenier\'ia Civil, Divisi\'on de Ingenier\'ia, Universidad de Guanajuato, Gto., M\'exico.}

\author{F. Antonio Horta-Rangel}\email{anthort@hotmail.com}\affiliation{Dpto. Ingenier\'ia Civil, Divisi\'on de Ingenier\'ia, Universidad de Guanajuato, Gto., M\'exico.}

\author{Joel Saavedra}\email{joel.saavedra@ucv.cl}\affiliation{Instituto de F\'isica, Pontificia Universidad Cat\'olica de Valpara\'iso,
Casilla 4950, Valpara\'iso, Chile.}

\date{\today}

\begin{abstract} This paper is aimed at a (mostly) pedagogical exposition of the derivation of the motion equations of certain modifications of general relativity. Here we derive in all detail the motion equations in the Brans-Dicke theory with the cubic self-interaction. This is a modification of the Brans-dicke theory by the addition of a term in the Lagrangian which is non-linear in the derivatives of the scalar field: it contains second-order derivatives. This is the basis of the so-called Brans-Dicke Galileon. We pay special attention to the variational principle and to the algebraic details of the derivation. It is shown how higher order derivatives of the fields appearing in the intermediate computations cancel out leading to second order motion equations. The reader will find useful tips for the derivation of the field equations of modifications of general relativity such as the scalar-tensor theories and $f(R)$ theories, by means of the (stationary action) variational principle. The content of this paper is specially recommended to those graduate and postgraduate students who are interested in the study of the mentioned modifications of general relativity.\end{abstract}


\maketitle

%%%%%%%%%%%%%%%%%%%%%%%%%%%%%%%%%%%
%%%%%%%%%%%%%%%%%%%%%%%%%%%%%%%%%%%

\section{Introduction}\label{intro}

In this paper we aim at a detailed derivation of the motion equations of Brans-Dicke (BD)-type theories with a term containing non-linear (higher order) derivatives of the scalar field, by means of the variational principle of least action or principle of stationary action \cite{feynman, book}. Here we systematically show the typical steps and the subtleties of the variational procedure, that, for obvious reasons, are usually not shown in standard research papers. The variational principle is well-known in many areas of science and its beauty and power has only strengthen with the time. For instance, several of its applications among which we may cite the already mentioned action principle, the Castigliano principle \cite{castig} and the Ritz method \cite{ritz}, are useful in different fields of knowledge such as theoretical physics (remarkably in the classical/quantum field theory), economy, ecology, biology, etc. 

The derivation of the field equations of general relativity (GR) is one of the best known applications of the action principle to those who have learned this subject \cite{landau, gr-book, carroll}. The derivation of the motion equations of modifications of GR such as the Brans-Dicke theory \cite{brans} is less known despite that scattered texts with specific details can be found. In the appendix C of Ref. \cite{fujii}, for instance, the field equations of gravity in the presence of non-minimal coupling with Lagrangian density ${\cal L}_\text{nmc}=\vphi R$ are derived. However, the derivation given in that reference is a bit complicated due to several subtle considerations made by the authors and, besides, a local Lorentz frame is invoked in order to simplify the computations. This means that the corresponding derivation could not seem straightforward to graduate and postgraduate researchers who are interested in the study of the related theories. 

Our goal in this paper will be to fill this gap: we shall derive the motion equations of the Brans-Dicke theory in all detail and without appealing, neither to clever considerations, nor to the choice of a specific frame. We shall be focusing in a variant of the BD theory, where a non-linear derivative self-interaction is added in the scalar sector. We choose this model because: i) BD theory represents the simplest modification of GR, yet the generalization of the procedure we shall explain, to more complex modifications such like the $f(R)$-theory \cite{fdr, odi-1, odi-2}, is straightforward,\footnote{For the derivation of the motion equations of $f(R)$-theory by means of the variational principle we recommend \cite{grg-2010}.} and ii) the introduction of the cubic self-interaction permits to show several subtleties of the application of the variational procedure when derivatives of the fields higher than the first one are implied (for instance, correct variation of the D'alembertian of the $\phi$-field), which deserve special consideration. 

We hope that our detailed exposition will help graduate and postgraduate students to learn how to derive the field equations of certain modifications of general relativity. Knowledge of the variational principle and the related procedure, gives us the freedom to look for new horizons, to seek for new ideas and models. One just slightly (and appropriately) modifies a known Lagrangian density, and the variational principle immediately yields the corresponding field equations. Usually, students ask to their advisers for working ideas, which means they want a set of master equations of the model subject of investigation, in order to be able to start the required study. Knowledge of the variational principle and of the associated variational procedure, will permit those students to obtain their own set of (modified) master equations, and this will contribute towards their independence during the research process.

As mentioned, here we focus in a modification of the Brans-Dicke theory where a piece of action containing derivatives of the Brans-Dicke field higher than the first one is considered. This modification is the basis of the so-called ``Brans-Dicke Galileon'' cosmological model,\footnote{The presently accepted cosmological paradigm asserts that the expansion of the universe is undergoing a period of accelerated expansion, which started recently in the cosmic history, at a redshift of about $z\approx 0.7$. The fundamental nature of this speedup of the expansion is not known, however, there are plenty of classical theoretical models that try to account for a plausible description of this stage of the cosmic evolution. Several of these models are based in Einstein's general relativity and need of an exotic component of the cosmic budget known as dark energy. Other models are based in modifications of GR. Among the latter ones the so called ``Galileons'' play an important role \cite{nicolis}. The name ``Galileon'' originates from the fact that, in the Minkowski spacetime, the field equations are invariant under the Galilean symmetry: $\der_\mu\phi\rightarrow\der_\mu\phi+b_\mu$. In these models, despite that the Lagrangian contains derivatives of order higher than one, the equation of motion for the scalar field -- properly the Galileon -- can remain of second order. This is essential because the higher-derivative theories contain extra degrees of freedom that are usually related to instabilities. In \cite{deffayet} the analysis of Ref. \cite{nicolis} was extended to the curved spacetime, and the most general covariant Lagrangian that keeps the field equations up to second-order was re-discovered (the original study is due to Horndeski \cite{horndeski}).} formerly studied in \cite{kazuya} (see also \cite{japan, chow, also}). The authors of the mentioned work demonstrated the existence of self-accelerating universe with no ghost-like instabilities on small scales if the Galileon is a BD scalar field $\phi$ with a cubic self-interaction term \cite{kazuya}. The action for this model is given by:

\bea S_\text{BD}^\text{cubic}=\int_{\cal M} d^4x\sqrt{|g|}\left[\phi R-\frac{\omega_\text{BD}}{\phi}(\der\phi)^2-2V(\phi)+\alpha^2\Box\phi\left(\frac{\der\phi}{\phi}\right)^2+2{\cal L}_m\right],\label{kazuya-action}\eea where $\sqrt{|g|}$ is a scalar density of weight $+1$ ($|g|$ is the absolute value of the determinant of the metric $g_{\mu\nu}$), so that $d^4x\sqrt{|g|}$ is an invariant measure, $R$ is the curvature scalar (also Ricci scalar), the BD scalar field $\phi$ stands as the Galileon field, $V(\phi)$ is the self-interaction potential for $\phi$, and $\omega_\text{BD}$ is a free constant called as the BD parameter. Besides, $\alpha^2$ -- the strength of the cubic self-interaction, is another free constant, $(\der\phi)^2\equiv g^{\mu\nu}\nabla_\mu\phi\nabla_\nu\phi=g^{\mu\nu}\der_\mu\phi\der_\nu\phi$, $\Box\phi\equiv g^{\mu\nu}\nabla_\mu\nabla_\nu\phi$, and ${\cal L}_m$ is the Lagrangian density of the matter degrees of freedom other than the Galileon itself. The cubic interaction $\propto\Box\phi(\der\phi)^2/\phi^2$, is the unique form of interactions at cubic order that keeps the field equation for the Galileon $\phi$ of second-order \cite{nicolis}. The existence of the self-accelerating universe requires a negative BD parameter $\omega_\text{BD}<0$, but, thanks to the non-linear term, small fluctuations around the solution are stable on small scales. General relativity is recovered at early times and on small scales by the cubic interaction via the Vainshtein mechanism.\footnote{This is a screening mechanism due to the non-linearity of the term containing the derivatives of the scalar field. For distances from the source (or at small cosmological scales) smaller than the Vainshtein radius $r_\text{V}$, which depends on the source and on the parameters of the theory, the gravitational effects of the scalar field are hidden via the non-linear effects so that, at distances $\ll r_\text{V}$ the theory is indistinguishable from general relativity \cite{vainsh-rev, chow}. The influence of the scalar field becomes important only at large scales, e. g. for cosmology.} At late time, gravity is strongly modified and the background cosmology shows a phantom-like behaviour \cite{kazuya}.

As said, our aim here is to show in all detail how to derive the motion equations by varying \eqref{kazuya-action} with respect to the fields: the metric $g_{\mu\nu}$ and the BD scalar field $\phi$ (properly the Galileon in the corresponding cosmological model), respectively. For simplicity of the derivation we shall split the action \eqref{kazuya-action} into two pieces: i) the standard Brans-Dicke theory's action,\footnote{The Brans-Dicke theory of gravity is a non-geometrical theory in the sense that the gravitational interactions are mediated both by the metric field and by the non-geometric BD scalar \cite{brans}. As a matter of fact the (inverse of the) Brans-Dicke scalar $\phi$ is associated with the gravitational coupling: $G\propto\phi^{-1}$, so that the point-dependent $\phi$-field (the Galileon in the model of \cite{kazuya}) modulates the local strength of the gravitational interactions. The spacetime curvature is only partly responsible for the gravitational phenomena. In the formal limit when $\omega_\text{BD}\rightarrow\infty$, the scalar field decouples from the gravity and we recover Einstein's GR from the BD theory. The Brans-Dicke action is equivalent to the one for the $f(R)$-theories when $\omega_\text{BD}=0$. In this case the different potentials $V$ correspond to different choices of the function $f(R)$ \cite{fdr, odi-1}.} 

\bea S_\text{BD}=\int_{\cal M} d^4x\sqrt{|g|}\left[\phi R-\frac{\omega_\text{BD}}{\phi}(\der\phi)^2-2V(\phi)+2{\cal L}_m\right],\label{bd-action}\eea and ii) the self-interacting (cubic) piece of action,

\bea S_\text{int}^\text{cubic}=\alpha^2\int_{\cal M} d^4x\sqrt{|g|}\,\Box\phi\left(\frac{\der\phi}{\phi}\right)^2.\label{cubic-action}\eea

The paper has been organized in the following way. In the next section (Sec. \ref{sec-2}) we concentrate in the Brans-Dicke theory without the cubic self-interaction. Variations with respect to the metric (Sec. \ref{subsec-2-1}) and with respect to the $\phi$-field (Sec. \ref{subsec-2-3}) are separately studied on purpose. In Sec. \ref{subsec-2-2}, variations of the $f(R)$-type action with respect to metric are considered as an additional illustration of the procedure exposed in Sec. \ref{subsec-2-1}. Variations of the non-linear self-interaction term containing the higher-order derivatives of the $\phi$-field, both with respect to the metric and with respect to the BD field, are investigated at length in Sec. \ref{sec-3}. In Sec. \ref{subsec-3-1}, particular focus is made in the variations with respect to the $\phi$-field. Several subtleties associated with the variational principle when applied to action terms with higher-order derivatives of the field, such as repeated use of the Stokes theorem and correct variation of the D'alembertian of the $\phi$-field, are shown. Some of these subtleties may go unnoticed even to experts. In Sec. \ref{sec-4} the results of the former sections are collected and the variational principle for the full action is shown. The corresponding motion equations are written. A generalization of the BD theory: the scalar-tensor theory in general, is studied in Sec. \ref{sec-stt-gen}, where we expose, again, the variational procedure in all details. Besides, in this section we apply the variational procedure to a cubic interaction term of the general form as well. In Sec. \ref{sec-tips} we leave to the readers, as an exercise, the derivation of the motion equations -- by means of the variational procedure -- from the coupled Galileon action \cite{ex}. Several helpful tips are provided to help the readers in the process of derivation of the motion equations. In Sec. \ref{sec-cosmo}, in order to put the procedure and the results exposed in the present paper in a context of interest for applications, we write the motion equations of the theory depicted by the action \eqref{kazuya-action}, in terms of the cosmological Friedmann-Robertson-Walker (FRW) metric with flat spatial sections. This is what is known as the BD-Galileon model \cite{kazuya}. A brief discussion of the main physical features of this cosmological model is also provided. Conclusions are given in Sec. \ref{sec-conclu}.

%%%%%%%%%%%%%%%%%%%%%%%%%%%%%%%%%%%%%%%%%%%%%%%%%%%%%%%%%%%%%%%%%%%%%%%
%%%%%%%%%%%%%%%%%%%%%%%%%%%%%%%%%%%%%%%%%%%%%%%%%%%%%%%%%%%%%%%%%%%%%%%

\section{Standard Brans-Dicke action and the variational principle}\label{sec-2}

In order to derive the motion equations of a given theory, by means of the variational principle, one have to perform independent variations with respect to the fields that enter the action. In the case of the Brans-Dicke theory, for instance, one needs to vary with respect to the metric $g_{\mu\nu}$ keeping the $\phi$-field fixed, and then one takes variations with respect to the $\phi$-field, with the metric field assumed fixed. In this paper, for pedagogical purposes, we shall separately discuss variations with respect to the metric (this subsection and Sec. \ref{subsec-2-1}) and with respect to the BD field (Sec. \ref{subsec-2-3}). 

Variation of the BD action \eqref{bd-action} with respect to the metric (the BD field assumed fixed), yields \cite{brans, fujii}

\bea \delta_g S_\text{BD}=\int_{\cal M} d^4x\sqrt{|g|}\delta g^{\mu\nu}\left[\phi G_{\mu\nu}-(\nabla_\mu\nabla_\nu-g_{\mu\nu}\Box)\phi-T^{(\phi)}_{\mu\nu}-T^{(m)}_{\mu\nu}\right],\label{var-bd-action}\eea where $G_{\mu\nu}\equiv R_{\mu\nu}-g_{\mu\nu} R/2$ is the Einstein's tensor, 

\bea T^{(\phi)}_{\mu\nu}\equiv\frac{\omega_\text{BD}}{\phi}\left[\der_\mu\phi\der_\nu\phi-\frac{1}{2}g_{\mu\nu}\left(\der\phi\right)^2\right]-g_{\mu\nu}V(\phi),\label{def-1}\eea is the stress-energy tensor of the $\phi$-field,\footnote{As a matter of fact the second and third terms under the integral in \eqref{var-bd-action}, collect the non-geometric part of the gravitational field.} and 

\bea T^{(m)}_{\mu\nu}:=-\frac{2}{\sqrt{|g|}}\frac{\delta\left(\sqrt{|g|}{\cal L}_m\right)}{\delta g^{\mu\nu}},\label{mat-set}\eea is the stress-energy tensor of the matter degrees of freedom. While computing the variation in \eqref{var-bd-action} we have taken into consideration the typical requirement that any fields variations, as well as variations of their first derivatives, vanish on the integration boundary $\der{\cal M}$. 

As a matter of fact, in order to have a well-posed variational principle, one should take into account the York-Gibbons-Hawking (YGH) boundary term $\propto\int_{\der\cal M}d^3x\sqrt{|h|}\phi K$, where $h_{\mu\nu}$ is the induced metric on the boundary $\der{\cal M}$ of the integration volume, and $K$ is its extrinsic curvature \cite{grg-2010, boundary}. Fortunately, we can work without considering the YGH term and yet we can obtain a well-posed theory. In this case the variational principle requires vanishing of the variation of the fields, as well as of the variation of their first derivatives, on the boundary of the integration volume. In the present paper we follow the latter formulation of the variational principle. For a detailed description of the former formulation when deriving the motion equations of the $f(R)$-type theories we recommend \cite{grg-2010}.

%%%%%%%%%%%%%%%%%%%%%%%%%%%%%%%%%%%%%%%%%%%%%%%%%%%%%%%%%%%%%%%%%%%%%%%%%%%%%%%%%%%%%%%%%%%%%%%%%%%%%%%%%

\subsection{The variational procedure in detail: variations with respect to the metric}\label{subsec-2-1}

Here we shall explain in as much details as possible the variational procedure that leads to \eqref{var-bd-action}. To start with let us to split the BD action \eqref{bd-action} into a pure gravitational part:

\bea S^\text{grav}_\text{BD}=\int_{\cal M} d^4x\sqrt{|g|}\left[\phi R-\frac{\omega_\text{BD}}{\phi}(\der\phi)^2-2V(\phi)\right],\label{pure-bd}\eea and a matter part\footnote{The factor of $2$ in the right-hand-side (RHS) of \eqref{matter-bd}, as well as in the matter Lagrangian in \eqref{kazuya-action}, is required if consider (as we do) that $\phi=1/8\pi G$, where $G$ is the point-dependent gravitational coupling. Notice, however, that if replace $1/16\pi G_N$ $\rightarrow\;\phi$, where $G_N$ is the Newton's constant, then the factor $2$ is not required, and $S_{(m)}=\int_{\cal M} d^4x\sqrt{|g|}{\cal L}_m$.} 

\bea S^{(m)}_\text{BD}=2\int_{\cal M} d^4x\sqrt{|g|}\,{\cal L}_m.\label{matter-bd}\eea Taking into account the definition of the stress-energy tensor in \eqref{mat-set}, the variation of this latter piece of action with respect to the $g_{\mu\nu}$ can be written as:

\bea \delta_g S^{(m)}_\text{BD}=2\int_{\cal M} d^4x\,\delta\left(\sqrt{|g|}\,{\cal L}_m\right)=-\int_{\cal M} d^4x\sqrt{|g|}\,\delta g^{\mu\nu}T^{(m)}_{\mu\nu}.\label{var-matter-bd}\eea 

Now let us to concentrate in the variation of the pure gravitational BD action \eqref{pure-bd}. Variation of \eqref{pure-bd} with respect to the metric, assuming that the $\phi$-field is fixed, reads:

\bea \delta_g S^\text{grav}_\text{BD}=\int_{\cal M} d^4x\left\{\delta\sqrt{|g|}\left[\phi R-\frac{\omega_\text{BD}}{\phi}(\der\phi)^2-2V(\phi)\right]+\sqrt{|g|}\left[\phi\delta R-\frac{\omega_\text{BD}}{\phi}\nabla_\mu\phi\nabla_\nu\phi\delta g^{\mu\nu}\right]\right\},\label{eq-x}\eea where it has been taken into account that $$\delta_g(\der\phi)^2=\delta_g\left(g^{\mu\nu}\nabla_\mu\phi\nabla_\nu\phi\right)=\delta g^{\mu\nu}\nabla_\mu\phi\nabla_\nu\phi.$$ Equation \eqref{eq-x} can be rewritten in a more feasible way by using the following expressions (see, for instance, Ref. \cite{landau, gr-book, carroll}):

\bea \delta\sqrt{|g|}=-\frac{1}{2}\sqrt{|g|}g_{\mu\nu}\delta g^{\mu\nu},\;\delta R=R_{\mu\nu}\delta g^{\mu\nu}+g^{\mu\nu}\delta R_{\mu\nu}.\label{useful-eqs}\eea Collecting the terms with the factor $\sqrt{|g|}\delta g^{\mu\nu}$, we get:

\bea \delta_g S^\text{grav}_\text{BD}=\int_{\cal M} d^4x\sqrt{|g|}\delta g^{\mu\nu}\left\{\phi G_{\mu\nu}-\frac{\omega_\text{BD}}{\phi}\left[\nabla_\mu\phi\nabla_\nu\phi-\frac{1}{2}g_{\mu\nu}(\der\phi)^2\right]+g_{\mu\nu}V(\phi)\right\}+\delta_g\bar S,\label{useful-1}\eea where

\bea \delta_g\bar S=\int_{\cal M}d^4x\sqrt{|g|}\,\phi g^{\mu\nu}\delta R_{\mu\nu}.\label{var-bar-action}\eea 

Let us now to focus in this latter piece of the variation. A useful expression is the (contraction of the) Palatini equation \cite{gr-book}

\bea \delta R_{\mu\nu}=\nabla_\lambda\left(\delta\Gamma^\lambda_{\mu\nu}\right)-\nabla_\nu\left(\delta\Gamma^\lambda_{\mu\lambda}\right),\label{palatini}\eea where, in our convention, the Christoffel symbols are given by:

\bea \Gamma^\alpha_{\mu\nu}=\frac{1}{2}g^{\alpha\lambda}\left(\der_\nu g_{\mu\lambda}+\der_\mu g_{\nu\lambda}-\der_\lambda g_{\mu\nu}\right),\label{christ}\eea while the components of the Ricci tensor are defined as it follows:

\bea R_{\mu\nu}=\der_\lambda\Gamma^\lambda_{\mu\nu}-\der_\nu\Gamma^\lambda_{\mu\lambda}+\Gamma^\lambda_{\mu\nu}\Gamma^\tau_{\lambda\tau}-\Gamma^\tau_{\mu\lambda}\Gamma^\lambda_{\tau\nu}.\label{ricci}\eea

Taking into account the Palatini equation, the variation \eqref{var-bar-action} can be written in the following way:

\bea \delta_g\bar S=\int_{\cal M}d^4x\sqrt{|g|}\,\phi g^{\mu\nu}\left[\nabla_\lambda\left(\delta\Gamma^\lambda_{\mu\nu}\right)-\nabla_\nu\left(\delta\Gamma^\lambda_{\mu\lambda}\right)\right].\label{var-bar-action-1}\eea We have

\bea &&\nabla_\lambda\left(\phi g^{\mu\nu}\delta\Gamma^\lambda_{\mu\nu}\right)=\nabla_\lambda\phi g^{\mu\nu}\delta\Gamma^\lambda_{\mu\nu}+\underline{\phi g^{\mu\nu}\nabla_\lambda\left(\delta\Gamma^\lambda_{\mu\nu}\right)},\nonumber\\
&&\nabla_\nu\left(\phi g^{\mu\nu}\delta\Gamma^\lambda_{\mu\lambda}\right)=\nabla_\nu\phi g^{\mu\nu}\delta\Gamma^\lambda_{\mu\lambda}+\underline{\phi g^{\mu\nu}\nabla_\nu\left(\delta\Gamma^\lambda_{\mu\lambda}\right)},\nonumber\eea where the divergence terms in the left-hand-side (LHS) of both equations above are usually called as boundary terms since these contribute an integral over the boundary of the integration volume. Actually, by virtue of the Stokes-Gauss-Ostrogradski theorem (hereafter, for short, just ``Gauss theorem'') $$\int_{\cal M} d^4x\sqrt{|g|}\nabla_\lambda\left(\phi g^{\mu\nu}\delta\Gamma^\lambda_{\mu\nu}\right)=\int_{\cal M} d^4x\,\der_\lambda\left(\sqrt{|g|}\phi g^{\mu\nu}\delta\Gamma^\lambda_{\mu\nu}\right)=\int_{\der\cal M} d\sigma_\lambda\,\phi g^{\mu\nu}\delta\Gamma^\lambda_{\mu\nu}=0,$$ and $$\int_{\cal M} d^4x\sqrt{|g|}\nabla_\nu\left(\phi g^{\mu\nu}\delta\Gamma^\lambda_{\mu\lambda}\right)=\int_{\cal M} d^4x\,\der_\nu\left(\sqrt{|g|}\phi g^{\mu\nu}\delta\Gamma^\lambda_{\mu\lambda}\right)=\int_{\der\cal M} d\sigma_\nu\,\phi g^{\mu\nu}\delta\Gamma^\lambda_{\mu\lambda}=0,$$ where $d\sigma^\mu$ are the components of the elementary hypersurface vector, and the vanishing of the surface integrals is due to the requirement of the stationary action principle (vanishing of variations of the metric and of its first derivatives on the integration boundary). This means that the boundary terms may be safely omitted. 

After the mentioned considerations the variation \eqref{var-bar-action-1} can be rewritten as:

\bea \delta_g\bar S=\int_{\cal M}d^4x\sqrt{|g|}\left[\nabla_\nu\phi g^{\mu\nu}\delta\Gamma^\lambda_{\mu\lambda}-\nabla_\lambda\phi g^{\mu\nu}\delta\Gamma^\lambda_{\mu\nu}\right]=\int_{\cal M}d^4x\sqrt{|g|}\,\nabla_\lambda\phi\left[g^{\mu\lambda}\delta\Gamma^\nu_{\mu\nu}-g^{\mu\nu}\delta\Gamma^\lambda_{\mu\nu}\right].\label{var-bar-action-2}\eea After a bit of algebra:\footnote{For details of this and related derivations we recommend the appendix A of Ref. \cite{grg-2010}. Particularly useful are the equations (A.7) and (A.8) in \cite{grg-2010}: $$\delta\Gamma^\lambda_{\mu\nu}=-\frac{1}{2}\left[g_{\nu\sigma}\nabla_\mu(\delta g^{\lambda\sigma})+g_{\mu\sigma}\nabla_\nu(\delta g^{\lambda\sigma})-g_{\mu\sigma}g_{\nu\tau}\nabla^\lambda(\delta g^{\sigma\tau})\right],\;\delta\Gamma^\lambda_{\mu\lambda}=-\frac{1}{2}g_{\lambda\sigma}\nabla_\mu(\delta g^{\lambda\sigma}),$$ respectively.}

\bea \nabla_\lambda\phi\left[g^{\mu\lambda}\delta\Gamma^\nu_{\mu\nu}-g^{\mu\nu}\delta\Gamma^\lambda_{\mu\nu}\right]=\nabla_\lambda\phi\left[\nabla_\mu(\delta g^{\mu\lambda})-g_{\mu\nu}\nabla^\lambda(\delta g^{\mu\nu})\right].\label{useful-2}\eea Hence:

\bea \delta_g\bar S=\int_{\cal M}d^4x\sqrt{|g|}\left[\nabla_\nu\phi\nabla_\mu\left(\delta g^{\mu\nu}\right)-\nabla_\lambda\phi g_{\mu\nu}\nabla^\lambda(\delta g^{\mu\nu})\right],\label{var-bar-action-3}\eea or, since

\bea &&\nabla_\mu\left(\nabla_\nu\phi \delta g^{\mu\nu}\right)=\nabla_\mu\nabla_\nu\phi \delta g^{\mu\nu}+\underline{\nabla_\nu\phi\nabla_\mu\left(\delta g^{\mu\nu}\right)},\nonumber\\
&&\nabla^\lambda\left(\nabla_\lambda\phi g_{\mu\nu}\delta g^{\mu\nu}\right)=\Box\phi g_{\mu\nu}\delta g^{\mu\nu}+\underline{\nabla_\lambda\phi g_{\mu\nu}\nabla^\lambda(\delta g^{\mu\nu})},\nonumber\eea where the LHS of both equations above amount to a divergence, whose integral over ${\cal M}$ vanishes by virtue of the Gauss theorem in conjunction with the requirements of the stationary action principle, then, finally:

\bea \delta_g\bar S=\int_{\cal M}d^4x\sqrt{|g|}\delta g^{\mu\nu}(g_{\mu\nu}\Box-\nabla_\mu\nabla_\nu)\phi.\label{var-bar-action-fin}\eea 

If substitute the variation $\delta_g\bar S$ from \eqref{var-bar-action-fin} back into \eqref{useful-1}, then, for the variation of the full action \eqref{bd-action}: $\delta S_\text{BD}$, one gets Eq. \eqref{var-bd-action}. By requiring vanishing of the variation in Eq. \eqref{var-bd-action}: $\delta_g S_\text{BD}=0,$ one obtains the Einstein-Brans-Dicke (EBD) field equations:

\bea G_{\mu\nu}=\frac{1}{\phi}\left[T^{(\phi)}_{\mu\nu}+T^{(m)}_{\mu\nu}\right]+\frac{1}{\phi}(\nabla_\mu\nabla_\nu-g_{\mu\nu}\Box)\phi,\label{ebd-feq}\eea or in more explicit form: 

\bea G_{\mu\nu}=\omega_\text{BD}\left[\frac{\der_\mu\phi}{\phi}\frac{\der_\nu\phi}{\phi}-\frac{1}{2}g_{\mu\nu}\left(\frac{\der\phi}{\phi}\right)^2\right]-g_{\mu\nu}\frac{V(\phi)}{\phi}+\frac{1}{\phi}(\nabla_\mu\nabla_\nu-g_{\mu\nu}\Box)\phi+\frac{T^{(m)}_{\mu\nu}}{\phi}.\label{ebd-feq'}\eea

%%%%%%%%%%%%%%%%%%%%%%%%%%%%%%%%%%%%%%%%%%%%%%%%%%%%%%%%%%%%

\subsection{The case of the $f(R)$-theory}\label{subsec-2-2}

The above explained procedure can be straightforwardly extrapolated to other modifications of general relativity such as the $f(R)$-theories \cite{fdr, odi-1, odi-2}. These theories come about by a straightforward generalization of the Lagrangian in the standard action of general relativity (the Einstein-Hilbert action): $$S_{f(R)}=\frac{1}{16\pi G_N}\int_{\cal M}d^4x\sqrt{|g|}\,f(R)\;\;\rightarrow\;\;S_\text{GR}=\frac{1}{16\pi G_N}\int_{\cal M}d^4x\sqrt{|g|}R,$$ where $f(R)$ can be any arbitrary algebraic function of the curvature scalar $R$. The $f(R)$ actions are simple enough, yet these are sufficiently general to encapsulate some of the basic characteristics of higher-order gravity, keeping the theory free from the fatal Ostrogradski instability.\footnote{Basically, what the Ostrogradski theorem states is that for Lagrangians which depend on derivatives of the fields of order higher than the first one, the associated Hamiltonian is unbounded from below, i. e., it is linearly unstable \cite{ostrogradski}. For an interesting discussion on this issue we recommend \cite{woodard}.} 

Take the $f(R)$ action and vary with respect to the metric field:

\bea \delta_g S_{f(R)}=\frac{1}{16\pi G_N}\int_{\cal M}d^4x\left[\delta\sqrt{|g|}\,f(R)+\sqrt{|g|}\delta f(R)\right]=\frac{1}{16\pi G_N}\int_{\cal M}d^4x\sqrt{|g|}\left[-\frac{1}{2}g_{\mu\nu} f(R) \delta g^{\mu\nu}+f_R \delta R\right],\label{fdr-action}\eea where $f_R\equiv df/dR$. Then, by following the same procedure explained above, one can write

\bea \delta_g S_{f(R)}=\frac{1}{16\pi G_N}\int_{\cal M}d^4x\sqrt{|g|}\delta g^{\mu\nu}\left[f_R R_{\mu\nu}-\frac{1}{2}g_{\mu\nu} f(R)\right]+\delta_g\hat S,\label{fdr-1}\eea where

\bea \delta_g\hat S=\frac{1}{16\pi G_N}\int_{\cal M}d^4x\sqrt{|g|}\,f_R g^{\mu\nu}\delta R_{\mu\nu}=\frac{1}{16\pi G_N}\int_{\cal M}d^4x\sqrt{|g|}\delta g^{\mu\nu}(g_{\mu\nu}\Box-\nabla_\mu\nabla_\nu)f_R.\label{fdr-2}\eea The first equality from the left in \eqref{fdr-2} is to be compared with \eqref{var-bar-action} in order to see how the last (right-hand) equality comes about. One finally gets:

\bea \delta_g S_{f(R)}=\frac{1}{16\pi G_N}\int_{\cal M}d^4x\sqrt{|g|}\delta g^{\mu\nu}\left[f_R R_{\mu\nu}-\frac{1}{2}g_{\mu\nu} f(R)-(\nabla_\mu\nabla_\nu-g_{\mu\nu}\Box)f_R\right].\label{var-fdr-action}\eea Besides, if include variation of the matter piece of action with respect to the metric:\footnote{We point out that variation of the matter piece of action with respect to the matter fields $\chi$ (the metric held fixed): $$\delta_\chi S_{(m)}=\int_{\cal M}d^4x\sqrt{|g|}\,\delta\chi\frac{\delta{\cal L}_m(\chi,\der\chi,g_{\mu\nu})}{\delta\chi},$$ and the requirement of vanishing of $\delta_\chi S_{(m)}=0$, yields to the conservation equations (see, for instance, Ref. \cite{landau}): $\nabla_\mu T^{\mu\nu}_{(m)}=0$.} $$\delta_g S_{(m)}=\int_{\cal M} d^4x\delta\left(\sqrt{|g|}{\cal L}_m\right)=-\frac{1}{2}\int_{\cal M} d^4x\sqrt{|g|}\delta g^{\mu\nu}T^{(m)}_{\mu\nu},$$ where we took into consideration \eqref{mat-set} and, due to the factor $1/16\pi G_N$ in \eqref{var-fdr-action}, the factor of $2$ in \eqref{matter-bd} has been omitted, then the requirement of the stationary action principle: $\delta_g S_\text{full}=\delta_g S_{f(R)}+\delta_g S_{(m)}=0$, leads to the motion equations of the $f(R)$ theory:

\bea f_R R_{\mu\nu}-\frac{1}{2}g_{\mu\nu} f(R)=(\nabla_\mu\nabla_\nu-g_{\mu\nu}\Box)f_R+8\pi G_N T^{(m)}_{\mu\nu}.\label{fdr-feq}\eea

%%%%%%%%%%%%%%%%%%%%%%%%%%%%%%%%%%%%%%%%%%%%%%%%%%%%%%%%%%%%%%%%%%%%%%%%%%%%%%%%%%%%%%%%%%%%%%%%%%%%%%%%%%%%%

\subsection{Variation with respect to the BD field: the Klein-Gordon-BD equation of motion}\label{subsec-2-3}

Variation with respect to the scalar field is, perhaps, the simplest part of the variational procedure. However, as we shall show soon, there are several subtleties even in this case. Variation of $S_\text{BD}$ with respect to the $\phi$-field (the metric $g_{\mu\nu}$ is held fixed) leads to:

\bea \delta_\phi S_\text{BD}=\int_{\cal M} d^4x\sqrt{|g|}\left[\delta\phi\,R-2\frac{\omega_\text{BD}}{\phi}\,\nabla^\mu\phi\nabla_\mu(\delta\phi)+\omega_\text{BD}\left(\frac{\der\phi}{\phi}\right)^2\delta\phi-2\der_\phi V\delta\phi\right],\label{eq-y}\eea where we have taken into consideration that 

\bea &&\delta_\phi\left[\phi^{-1}(\der\phi)^2\right]=-\phi^{-2}(\der\phi)^2\delta\phi+\delta_\phi\left(\nabla^\mu\phi\nabla_\mu\phi\right)\nonumber\\
&&\;\;\;\;\;\;\;\;\;\;\;\;\;\;\;\;\;\;\;\;\;\;=\left(\frac{\der\phi}{\phi}\right)^2\delta\phi+\nabla^\mu(\delta\phi)\nabla_\mu\phi+\nabla^\mu\phi\nabla_\mu(\delta\phi)=\left(\frac{\der\phi}{\phi}\right)^2\delta\phi+2\nabla^\mu\phi\nabla_\mu(\delta\phi).\nonumber\eea Besides, if realize that

\bea \nabla_\mu\left(\frac{\nabla^\mu\phi}{\phi}\,\delta\phi\right)=\nabla_\mu V^\mu=\frac{\Box\phi}{\phi}\delta\phi-\left(\frac{\der\phi}{\phi}\right)^2\delta\phi+\frac{\nabla^\mu\phi}{\phi}\nabla_\mu(\delta\phi),\nonumber\eea where we have introduced the vector $V^\mu\equiv(\nabla^\mu\phi/\phi)\delta\phi,$ then the equation \eqref{eq-y} can be written as

\bea \delta_\phi S_\text{BD}=\int_{\cal M} d^4x\sqrt{|g|}\delta\phi\left[R+2\omega_\text{BD}\frac{\Box\phi}{\phi}-\omega_\text{BD}\left(\frac{\der\phi}{\phi}\right)^2-2\der_\phi V\right],\label{var-phi-bd-action}\eea where, by virtue of the Gauss theorem and of the requirements of the variational procedure undertaken here, the boundary term: $$\int_{\cal M} d^4x\sqrt{|g|}\,\nabla_\mu V^\mu=\int_{\cal M} d^4x\,\der_\mu(\sqrt{|g|}\,V^\mu)=\int_{\der\cal M} d\sigma_\mu V^\mu=0,$$ vanishes.

The Klein-Gordon-Brans-Dicke (KGBD) equation for the BD field is obtained by requiring that $\delta_\phi S_\text{BD}=0$:

\bea 2\omega_\text{BD}\frac{\Box\phi}{\phi}-\omega_\text{BD}\left(\frac{\der\phi}{\phi}\right)^2+R=2\der_\phi V,\label{kgbd-eq}\eea or, if substitute the trace of Eq. \eqref{ebd-feq}:

\bea R=\omega_\text{BD}\left(\frac{\der\phi}{\phi}\right)^2+4\frac{V}{\phi}+3\frac{\Box\phi}{\phi}-\frac{T^{(m)}}{\phi},\label{r-trace}\eea back into \eqref{kgbd-eq}, the latter equation can be finally written in the known form \cite{brans}:

\bea (3+2\omega_\text{BD})\Box\phi=2\phi\der_\phi V-4V+T^{(m)}.\label{kg-bd-eq}\eea

The equations \eqref{ebd-feq'} and \eqref{kg-bd-eq} are the standard motion equations of the Brans-Dicke theory with the self-interacting scalar field.

%%%%%%%%%%%%%%%%%%%%%%%%%%%%%%%%%%%%%%%%%%%%%%%%%%%%%%%%%%%%%%%%%%%
%%%%%%%%%%%%%%%%%%%%%%%%%%%%%%%%%%%%%%%%%%%%%%%%%%%%%%%%%%%%%%%%%%%

\section{Variation of the cubic self-interaction term}\label{sec-3}

It is necessary to vary the piece of action \eqref{cubic-action} -- the one containing the cubic self-interaction -- with respect to the fields $g_{\mu\nu}$ and $\phi$ respectively, in order to obtain the corresponding corrections to the standard BD equations of motion \eqref{ebd-feq} and \eqref{kg-bd-eq}. Variation of this piece of action contains derivatives of the 3rd order which lead (in principle) to derivatives of the 3rd order in the motion equations. However, it is well-known that for Lagrangians of the Horndeski type such as \cite{nicolis, deffayet, horndeski}: ${\cal L}=\sqrt{|g|}\Box\phi(\der\phi/\phi)^2$, the corresponding motion equations are second-order, so that the theory is free from the fatal Ostrogradski instability. In our detailed exposition below we shall see how the terms with 3rd order derivatives on $\phi$, appearing in the intermediate computations, cancel out thus leading to 2nd order motion equations.

%%%%%%%%%%%%%%%%%%%%%%%%%%%%%%%%%%%%%%%%%%%%%%%%%%%%%%%%%%%%%%%%%%%%%

\subsection{Contribution of the cubic interaction towards the EBD motion equations}

Variation of \eqref{cubic-action} with respect to the metric (the $\phi$-field held fixed): 

\bea &&\delta_g S_\text{int}^\text{cubic}=\alpha^2\int_{\cal M} d^4x\left\{\delta\sqrt{|g|}\Box\phi\left(\frac{\der\phi}{\phi}\right)^2+\sqrt{|g|}\Box\phi\frac{\nabla_\mu\phi}{\phi}\frac{\nabla_\nu\phi}{\phi}\delta g^{\mu\nu}\right.\nonumber\\
&&\left.\;\;\;\;\;\;\;\;\;\;\;\;\;\;\;\;\;\;\;\;\;\;\;\;\;\;\;\;\;\;\;\;\;\;\;\;+\sqrt{|g|}\left(\frac{\der\phi}{\phi}\right)^2\left[\delta g^{\mu\nu}\nabla_\mu\nabla_\nu\phi+\nabla_\nu\phi\nabla_\mu(\delta g^{\nu\mu})-\frac{1}{2}g_{\mu\nu}\nabla_\lambda\phi\nabla^\lambda(\delta g^{\mu\nu})\right]\right\},\label{var-3bic}\eea where we have taken into account the fact that $$\delta_g\Box\phi=\delta_g\left[g^{\mu\nu}\nabla_\mu\nabla_\nu\phi\right]=\delta g^{\mu\nu}\nabla_\mu\nabla_\nu\phi-g^{\mu\nu}\delta\Gamma^\lambda_{\mu\nu}\nabla_\lambda\phi,$$ which, after equation (A.7) of \cite{grg-2010}: $$-g^{\mu\nu}\delta\Gamma^\lambda_{\mu\nu}\nabla_\lambda\phi=\nabla_\lambda\phi\nabla_\mu(\delta g^{\lambda\mu})-\frac{1}{2}g_{\mu\nu}\nabla_\lambda\phi\nabla^\lambda(\delta g^{\mu\nu}),$$ yields that: 

\bea \delta_g\left[\Box\phi\left(\frac{\der\phi}{\phi}\right)^2\right]=\left(\frac{\der\phi}{\phi}\right)^2\left[\delta g^{\mu\nu}\nabla_\mu\nabla_\nu\phi+\nabla_\lambda\phi\nabla_\mu(\delta g^{\lambda\mu})-\frac{1}{2}g_{\mu\nu}\nabla_\lambda\phi\nabla^\lambda(\delta g^{\mu\nu})\right]+\Box\phi\frac{\nabla_\mu\phi}{\phi}\frac{\nabla_\nu\phi}{\phi}\delta g^{\mu\nu}.\nonumber\eea Besides,

\bea &&\nabla_\mu\left[\delta g^{\mu\nu}\nabla_\nu\phi\left(\frac{\der\phi}{\phi}\right)^2\right]=\delta g^{\mu\nu}\nabla_\mu\left[\nabla_\nu\phi\left(\frac{\der\phi}{\phi}\right)^2\right]+\underline{\nabla_\mu(\delta g^{\mu\nu})\nabla_\nu\phi\left(\frac{\der\phi}{\phi}\right)^2},\nonumber\\
&&\nabla^\lambda\left[g_{\mu\nu}\delta g^{\mu\nu}\nabla_\lambda\phi\left(\frac{\der\phi}{\phi}\right)^2\right]=g_{\mu\nu}\delta g^{\mu\nu}\nabla^\lambda\left[\nabla_\lambda\phi\left(\frac{\der\phi}{\phi}\right)^2\right]+\underline{g_{\mu\nu}\nabla^\lambda(\delta g^{\mu\nu})\nabla_\lambda\phi\left(\frac{\der\phi}{\phi}\right)^2},\label{bound}\eea where the boundary terms $\propto\nabla(\delta g\nabla\phi...)$ in the LHS of both equations in \eqref{bound} may be safely omitted, hence 

\bea &&\underline{\left(\frac{\der\phi}{\phi}\right)^2\left[\nabla_\mu(\delta g^{\mu\nu})\nabla_\nu\phi-\frac{1}{2}g_{\mu\nu}\nabla^\lambda(\delta g^{\mu\nu})\nabla_\lambda\phi\right]}=-\delta g^{\mu\nu}\left\{\nabla_\mu\left[\nabla_\nu\phi\left(\frac{\der\phi}{\phi}\right)^2\right]-\frac{1}{2}g_{\mu\nu}\nabla^\lambda\left[\nabla_\lambda\phi\left(\frac{\der\phi}{\phi}\right)^2\right]\right\}\nonumber\\
&&\;\;\;\;\;\;\;\;\;\;\;\;\;\;\;\;\;\;\;\;\;\;\;\;\;\;\;=-\delta g^{\mu\nu}\left[\nabla_\mu\nabla_\nu\phi\left(\frac{\der\phi}{\phi}\right)^2-\frac{1}{2}g_{\mu\nu}\Box\phi\left(\frac{\der\phi}{\phi}\right)^2+\nabla_\nu\phi\nabla_\mu\left(\frac{\der\phi}{\phi}\right)^2-\frac{1}{2}g_{\mu\nu}\nabla_\lambda\phi\nabla^\lambda\left(\frac{\der\phi}{\phi}\right)^2\right].\nonumber\eea After substituting the underlined terms above into \eqref{var-3bic}, the latter equation can be written as:

\bea \delta_g S_\text{int}^\text{cubic}=\alpha^2\int_{\cal M} d^4x\sqrt{|g|}\delta g^{\mu\nu}\left[\Box\phi\frac{\nabla_\mu\phi}{\phi}\frac{\nabla_\nu\phi}{\phi}-\nabla_{(\mu}\phi\nabla_{\nu)}\left(\frac{\der\phi}{\phi}\right)^2+\frac{1}{2}g_{\mu\nu}\nabla_\lambda\phi\nabla^\lambda\left(\frac{\der\phi}{\phi}\right)^2\right],\label{var-g-cubic}\eea where we have to realize that, due to the symmetric property of the metric tensor: 

\bea \delta g^{\mu\nu}\nabla_\nu\phi\nabla_\mu\left(\frac{\der\phi}{\phi}\right)^2=\frac{1}{2}\delta g^{\mu\nu}\left[\nabla_\mu\phi\nabla_\nu\left(\frac{\der\phi}{\phi}\right)^2+\nabla_\nu\phi\nabla_\mu\left(\frac{\der\phi}{\phi}\right)^2\right]=\delta g^{\mu\nu}\nabla_{(\mu}\phi\nabla_{\nu)}\left(\frac{\der\phi}{\phi}\right)^2.\label{step}\eea This is a very subtle fact that, if not taken into account, may lead to a misleading result with a non-symmetric term in the (by themselves symmetric) Einstein's equations. 

The variation of the full action $S_\text{BD}^\text{cubic}$ with respect to the metric is given by:

\bea &&\delta_g S_\text{BD}^\text{cubic}=\delta_g S_\text{BD}+\delta_g S_\text{int}^\text{cubic}=\int_{\cal M} d^4x\sqrt{|g|}\delta g^{\mu\nu}\left[\phi G_{\mu\nu}-(\nabla_\mu\nabla_\nu-g_{\mu\nu}\Box)\phi-T^{(\phi)}_{\mu\nu}-T^{(\alpha^2)}_{\mu\nu}-T^{(m)}_{\mu\nu}\right],\label{var-g-full}\eea where $T^{(\phi)}_{\mu\nu}$ and $T^{(m)}_{\mu\nu}$ are given by \eqref{def-1} and \eqref{mat-set}, respectively, and we have associated a ``stress-energy tensor'' with the cubic self-interaction:

\bea T^{(\alpha^2)}_{\mu\nu}=-\alpha^2\left[\Box\phi\frac{\nabla_\mu\phi}{\phi}\frac{\nabla_\nu\phi}{\phi}-\nabla_{(\mu}\phi\nabla_{\nu)}\left(\frac{\der\phi}{\phi}\right)^2+\frac{1}{2}g_{\mu\nu}\nabla_\lambda\phi\nabla^\lambda\left(\frac{\der\phi}{\phi}\right)^2\right].\label{cub-sec}\eea

%%%%%%%%%%%%%%%%%%%%%%%%%%%%%%%%%%%%%%%%%%%%%%%%%%%%%%%%%%%%%%%%%%%%%%%%%%%%%%%%%%%%%%%%%%%%%%%%%%%%%%%%%%%

\subsection{Contribution of the cubic self-interaction towards the KGBD motion equation}\label{subsec-3-1}

The variation of \eqref{cubic-action} with respect to the BD field, keeping the metric $g_{\mu\nu}$ fixed, is also complex due to the higher derivatives of $\phi$, so that here we will perform it in all details as well. Straightforward variation with respect to $\phi$ yields

\bea \delta_\phi S_\text{int}^\text{cubic}=\alpha^2\int_{\cal M} d^4x\sqrt{|g|}\left[\Box(\delta\phi)\left(\frac{\der\phi}{\phi}\right)^2+2\frac{\Box\phi}{\phi}\frac{\nabla_\mu\phi}{\phi}\nabla^\mu(\delta\phi)-2\frac{\Box\phi}{\phi}\left(\frac{\der\phi}{\phi}\right)^2\delta\phi\right].\label{1}\eea In order to take the variation $\delta\phi$ out of the operators $\Box$ and $\nabla_\mu$, we resort to the following equations:

\bea &&\nabla_\mu\left[\nabla^\mu(\delta\phi)\left(\frac{\der\phi}{\phi}\right)^2\right]=\underline{\Box(\delta\phi)\left(\frac{\der\phi}{\phi}\right)^2}+\nabla^\mu(\delta\phi)\nabla_\mu\left(\frac{\der\phi}{\phi}\right)^2,\nonumber\\
&&\nabla^\mu\left(\frac{\Box\phi}{\phi}\frac{\nabla_\mu\phi}{\phi}\,\delta\phi\right)=\frac{\nabla^\mu(\Box\phi)}{\phi}\frac{\nabla_\mu\phi}{\phi}\,\delta\phi-2\frac{\Box\phi}{\phi}\left(\frac{\der\phi}{\phi}\right)^2\delta\phi+\left(\frac{\Box\phi}{\phi}\right)^2\delta\phi+\underline{\frac{\Box\phi}{\phi}\frac{\nabla_\mu\phi}{\phi}\nabla^\mu(\delta\phi)}.\label{2}\eea Notice that the terms in the LHS of the above equations amount to a divergence: $$\nabla_\mu\left[\nabla^\mu(\delta\phi)\left(\frac{\der\phi}{\phi}\right)^2\right]=\nabla_\mu V_1^\mu,\;\nabla_\mu\left(\frac{\Box\phi}{\phi}\frac{\nabla^\mu\phi}{\phi}\,\delta\phi\right)=\nabla_\mu V_2^\mu,$$ where $$V_1^\mu=\nabla^\mu(\delta\phi)\left(\frac{\der\phi}{\phi}\right)^2,\;V_2^\mu=\frac{\Box\phi}{\phi}\frac{\nabla^\mu\phi}{\phi}\,\delta\phi.$$ Then, we recall that by virtue of the Gauss theorem: $$\int_{\cal M} d^4x\sqrt{|g|}\,\nabla_\mu V_{1,2}^\mu=\int_{\cal M} d^4x\,\der_\mu (\sqrt{|g|}V_{1,2}^\mu)=\int_{\der\cal M}d\sigma_\mu V_{1,2}^\mu,$$ where the surface integrals vanish due to the requirement of the stationary action principle. This means that, to all purposes, the LHS of both equations in \eqref{2} can be safely set to zero. We can use the resulting equations to substitute the terms with the $\Box(\delta\phi)$ and $\nabla_\mu(\delta\phi)$ (underlined terms in \eqref{2}), back into \eqref{1}. We get:

\bea \delta_\phi S_\text{int}^\text{cubic}=\alpha^2\int_{\cal M} d^4x\sqrt{|g|}\left[-\nabla_\mu\left(\frac{\der\phi}{\phi}\right)^2\nabla^\mu(\delta\phi)-2\frac{\nabla_\mu(\Box\phi)}{\phi}\frac{\nabla^\mu\phi}{\phi}\,\delta\phi-2\left(\frac{\Box\phi}{\phi}\right)^2\delta\phi+2\frac{\Box\phi}{\phi}\left(\frac{\der\phi}{\phi}\right)^2\delta\phi\right].\label{3}\eea 

At this intermediate step let us note that, thanks to the factor $\propto\nabla_\mu(\Box\phi)$, the second term under the integral in \eqref{3} shows derivatives of the 3rd order. This would not bother us since, as said, the Lagrangian density under integral in \eqref{cubic-action} belongs in the class of Horndeski-type Lagrangians \cite{nicolis, deffayet, horndeski} and, hence, it would not lead to terms with derivatives of order higher than 2 in the field equations. In the following computations we shall see how this term with higher-order derivatives is compensated by a similar term with opposite sign coming from integrating by parts the first term under integral in \eqref{3}. Actually, let us to take the variation $\delta\phi$ out of the operator in $\nabla^\mu(\delta\phi)$ (the mentioned first term under the integral in \eqref{3}). For this purpose we use the following equation: $$\nabla^\mu\left[\nabla_\mu\left(\frac{\der\phi}{\phi}\right)^2\delta\phi\right]=\Box\left(\frac{\der\phi}{\phi}\right)^2\delta\phi+\underline{\nabla_\mu\left(\frac{\der\phi}{\phi}\right)^2\nabla^\mu(\delta\phi)},$$ where the LHS is a divergence and, as in the former cases, can be safely taken vanishing, so that: 

\bea &&-\nabla_\mu\left(\frac{\der\phi}{\phi}\right)^2\nabla^\mu(\delta\phi)=\Box\left(\frac{\der\phi}{\phi}\right)^2\delta\phi=\delta\phi\left[2\frac{\nabla_\mu(\Box\phi)}{\phi}\frac{\nabla^\mu\phi}{\phi}+2R_{\mu\nu}\frac{\nabla^\mu\phi}{\phi}\frac{\nabla^\nu\phi}{\phi}+2\frac{\nabla^\mu\nabla^\nu\phi}{\phi}\frac{\nabla_\mu\nabla_\nu\phi}{\phi}\right.\nonumber\\
&&\left.\;\;\;\;\;\;\;\;\;\;\;\;\;\;\;\;\;\;\;\;\;\;\;\;\;\;\;\;\;\;\;\;\;\;\;\;\;\;\;\;\;\;\;\;\;\;\;\;\;\;\;\;\;\;\;\;\;\;\;\;\;\;\;\;\;\;\;-8\frac{\nabla^\mu\nabla^\nu\phi}{\phi}\frac{\nabla_\mu\phi}{\phi}\frac{\nabla_\nu\phi}{\phi}-2\frac{\Box\phi}{\phi}\left(\frac{\der\phi}{\phi}\right)^2+6\left(\frac{\der\phi}{\phi}\right)^4\right].\label{4}\eea The right-hand equality above is obtained since:

\bea &&\Box\left(\frac{\der\phi}{\phi}\right)^2=\nabla^\mu\nabla_\mu\left[(\nabla^\nu\phi\nabla_\nu\phi)\phi^{-2}\right]=\nabla^\mu\left[2\nabla_\nu\phi\nabla_\mu(\nabla^\nu\phi)\phi^{-2}-2(\der\phi)^2\nabla_\mu\phi\,\phi^{-3}\right]\nonumber\\
&&\;\;\;\;\;\;\;\;\;\;\;\;\;\;\;\;=2\frac{\nabla^\mu\nabla^\nu\phi}{\phi}\frac{\nabla_\mu\nabla_\nu\phi}{\phi}+2\frac{\nabla^\nu\phi}{\phi}\frac{\Box(\nabla_\nu\phi)}{\phi}-4\frac{\nabla_\nu\phi}{\phi}\frac{\nabla_\mu\nabla^\nu\phi}{\phi}\frac{\nabla^\mu\phi}{\phi}\nonumber\\
&&\;\;\;\;\;\;\;\;\;\;\;\;\;\;\;\;\;\;\,\;-4\frac{\nabla_\nu\phi}{\phi}\frac{\nabla^\mu\nabla^\nu\phi}{\phi}\frac{\nabla_\mu\phi}{\phi}-2\frac{\nabla^\nu\phi}{\phi}\frac{\nabla_\nu\phi}{\phi}\frac{\Box\phi}{\phi}+6\frac{\nabla^\nu\phi}{\phi}\frac{\nabla_\nu\phi}{\phi}\frac{\nabla_\mu\phi}{\phi}\frac{\nabla^\mu\phi}{\phi}\nonumber\\
&&\;\;\;\;\;\;\;\;\;\;\;\;\;\;\;\;=2\frac{\nabla^\mu\nabla^\nu\phi}{\phi}\frac{\nabla_\mu\nabla_\nu\phi}{\phi}+2\frac{\nabla_\mu(\Box\phi)}{\phi}\frac{\nabla^\mu\phi}{\phi}+2R_{\mu\nu}\frac{\nabla^\mu\phi}{\phi}\frac{\nabla^\nu\phi}{\phi}\nonumber\\
&&\;\;\;\;\;\;\;\;\;\;\;\;\;\;\;\;\;\;\;\;\;-8\frac{\nabla^\mu\nabla^\nu\phi}{\phi}\frac{\nabla_\mu\phi}{\phi}\frac{\nabla_\nu\phi}{\phi}-2\frac{\Box\phi}{\phi}\left(\frac{\der\phi}{\phi}\right)^2+6\left(\frac{\der\phi}{\phi}\right)^4,\label{xx}\eea where, given that the metric is convariantly constant $\nabla_\lambda g_{\mu\nu}=0$, then $\nabla^\mu\nabla_\nu\phi\nabla_\mu\nabla^\nu\phi=\nabla^\mu\nabla^\nu\phi\nabla_\mu\nabla_\nu\phi$. Besides, in the last equality in \eqref{xx}, we have used the non-commutativity property of the covariant derivatives of a vector field: $$\nabla_\mu\nabla_\nu V^\alpha-\nabla_\nu\nabla_\mu V^\alpha=R^\alpha_{\;\;\lambda\mu\nu}V^\lambda\;\;\Rightarrow\;\;\nabla_\mu\nabla_\nu V^\mu-\nabla_\nu\nabla_\mu V^\mu=R_{\mu\nu}V^\mu,$$ in order to obtain the following useful relationship ($\nabla^\mu\phi\rightarrow V^\mu$): 

\bea \Box(\nabla_\mu\phi)-\nabla_\mu(\Box\phi)=R_{\mu\nu}\nabla^\nu\phi.\label{nc-useful}\eea If substitute the term $\propto\nabla^\mu(\delta\phi)$ from \eqref{4} into \eqref{3}, one finally gets:

\bea \delta_\phi S_\text{int}^\text{cubic}=2\alpha^2\int_{\cal M} d^4x\sqrt{|g|}\,\delta\phi\left\{\left[R_{\mu\nu}-4\frac{\nabla_\mu\nabla_\nu\phi}{\phi}\right]\frac{\nabla^\mu\phi}{\phi}\frac{\nabla^\nu\phi}{\phi}+\frac{\nabla^\mu\nabla^\nu\phi}{\phi}\frac{\nabla_\mu\nabla_\nu\phi}{\phi}+3\left(\frac{\der\phi}{\phi}\right)^4-\left(\frac{\Box\phi}{\phi}\right)^2\right\}.\label{var-phi-cubic}\eea 

Notice that, as promised, the term with the 3rd order derivatives $\propto\nabla_\mu(\Box\phi)$ in the last equation in \eqref{xx} compensates the similar (second) term in the integral in \eqref{3} with the opposite sign, so that in \eqref{var-phi-cubic} there are no terms with derivatives of the scalar field of order higher than the 2nd. Put in different words: if take a look at \eqref{nc-useful}, what we have done is to trade the higher derivatives of the $\phi$-field by the term containing the Ricci tensor.

After this the variation of the full action \eqref{kazuya-action} with respect to the BD field $\delta_\phi S_\text{BD}^\text{cubic}=\delta_\phi S_\text{BD}+\delta_\phi S_\text{int}^\text{cubic}$, reads:

\bea &&\delta_\phi S_\text{BD}^\text{cubic}=\int_{\cal M} d^4x\sqrt{|g|}\delta\phi\left[R+2\omega_\text{BD}\frac{\Box\phi}{\phi}-\omega_\text{BD}\left(\frac{\der\phi}{\phi}\right)^2-2\der_\phi V+2\alpha^2\left[R_{\mu\nu}-4\frac{\nabla_\mu\nabla_\nu\phi}{\phi}\right]\frac{\nabla^\mu\phi}{\phi}\frac{\nabla^\nu\phi}{\phi}\right.\nonumber\\
&&\left.\;\;\;\;\;\;\;\;\;\;\;\;\;\;\;\;\;\;\;\;\;\;\;\;\;\;\;\;\;\;\;\;\;\;\;\;\;\;\;\;\;\;\;\;\;\;\;\;\;\;\;\;\;\;\;\;\;\;\;\;\;\;\;\;\;\;\;\;\;\;\;\;\;\;\;\;\;+2\alpha^2\frac{\nabla^\mu\nabla^\nu\phi}{\phi}\frac{\nabla_\mu\nabla_\nu\phi}{\phi}+6\alpha^2\left(\frac{\der\phi}{\phi}\right)^4-2\alpha^2\left(\frac{\Box\phi}{\phi}\right)^2\right].\label{var-phi-full}\eea 

Due to the higher-order derivatives of the cubic self-interaction term the path from \eqref{1} to \eqref{var-phi-cubic} has been long and a bit complicated, involving repeated application of the Gauss theorem.

%%%%%%%%%%%%%%%%%%%%%%%%%%%%%%%%%%%%%%%%%%%%%%%%%%%%%%%%%%%%%%%%%%%%%%%%%%%%%%%%%%%%%%%%%%%
%%%%%%%%%%%%%%%%%%%%%%%%%%%%%%%%%%%%%%%%%%%%%%%%%%%%%%%%%%%%%%%%%%%%%%%%%%%%%%%%%%%%%%%%%%%

\section{Full variational principle and Einstein-Brans-Dicke motion equations}\label{sec-4}

Collecting the above results one sees that requiring vanishing of the variation of the action \eqref{kazuya-action} with respect to the metric $\delta_g S_\text{BD}^\text{cubic}=\delta_g S_\text{BD}+\delta_g S_\text{int}^\text{cubic}=0$, in Eq. \eqref{var-g-full}, leads to the following Einstein-BD equations: 

\bea G_{\mu\nu}=\frac{1}{\phi}\left[T^{(\phi)}_{\mu\nu}+T^{(\alpha^2)}_{\mu\nu}+T^{(m)}_{\mu\nu}\right]+\frac{1}{\phi}(\nabla_\mu\nabla_\nu-g_{\mu\nu}\Box)\phi,\label{e-bd-eq'}\eea where $T^{(\phi)}_{\mu\nu}$ and $T^{(\alpha^2)}_{\mu\nu}$, are given by equations \eqref{def-1} and \eqref{cub-sec} respectively, while requiring vanishing of $\delta_\phi S_\text{BD}^\text{cubic}=\delta_\phi S_\text{BD}+\delta_\phi S_\text{int}^\text{cubic}=0$, in \eqref{var-phi-full}, leads to: 

\bea &&2\omega_\text{BD}\frac{\Box\phi}{\phi}-2\alpha^2\left(\frac{\Box\phi}{\phi}\right)^2-\omega_\text{BD}\left(\frac{\der\phi}{\phi}\right)^2+6\alpha^2\left(\frac{\der\phi}{\phi}\right)^4+2\alpha^2\frac{\nabla^\mu\nabla^\nu\phi}{\phi}\frac{\nabla_\mu\nabla_\nu\phi}{\phi}\nonumber\\
&&\;\;\;\;\;\;\;\;\;\;\;\;\;\;\;\;\;\;\;\;\;\;\;\;\;\;\;\;\;\;\;\;\;\;\;\;\;\;\;\;\;\;\;\;\;\;\;\;\;\;\;\;\;\;\;\;\;\;\;\;\;\;+2\alpha^2\left[R_{\mu\nu}-4\frac{\nabla_\mu\nabla_\nu\phi}{\phi}\right]\frac{\nabla^\mu\phi}{\phi}\frac{\nabla^\nu\phi}{\phi}+R=2\der_\phi V,\label{5}\eea or, if take into account the trace of Eq. \eqref{e-bd-eq'} and substitute the obtained expression of the curvature scalar: $$R=\omega_\text{BD}\left(\frac{\der\phi}{\phi}\right)^2+4\frac{V}{\phi}+3\frac{\Box\phi}{\phi}+\alpha^2\left[\frac{\Box\phi}{\phi}\left(\frac{\der\phi}{\phi}\right)^2+\frac{\nabla_\lambda\phi}{\phi}\nabla^\lambda\left(\frac{\der\phi}{\phi}\right)^2\right]-\frac{T^{(m)}}{\phi},$$ into \eqref{5}, the resulting KGBD equation reads:

\bea &&\left[3+2\omega_\text{BD}+\alpha^2\left(\frac{\der\phi}{\phi}\right)^2\right]\frac{\Box\phi}{\phi}-2\alpha^2\left(\frac{\Box\phi}{\phi}\right)^2+6\alpha^2\left(\frac{\der\phi}{\phi}\right)^4+\alpha^2\frac{\nabla_\lambda\phi}{\phi}\nabla^\lambda\left(\frac{\der\phi}{\phi}\right)^2\nonumber\\
&&\;\;\;\;\;\;\;\;\;\;\;\;\;\;\;\;\;\;\;\;\;\;\;\;\;\;\;\;\;\;\;\;\;\;+2\alpha^2\frac{\nabla^\mu\nabla^\nu\phi}{\phi}\frac{\nabla_\mu\nabla_\nu\phi}{\phi}+2\alpha^2\left[R_{\mu\nu}-4\frac{\nabla_\mu\nabla_\nu\phi}{\phi}\right]\frac{\nabla^\mu\phi}{\phi}\frac{\nabla^\nu\phi}{\phi}=2\frac{\phi\der_\phi V-2V}{\phi}+\frac{T^{(m)}}{\phi}.\label{kg-bd-eq'}\eea 

The (master) motion equations of the Brans-Dicke Galileon with the cubic self-interaction \cite{kazuya} are the equations \eqref{e-bd-eq'} and \eqref{kg-bd-eq'}.

Notice that in the formal limit $\alpha^2\rightarrow 0$, the above equations reduce to the standard equations of the BD theory of gravity with the potential. Further, the constant field limit $\phi=\phi_0$, leaves us with the standard Einstein's GR with a cosmological constant $\Lambda=V(\phi_0)/\phi_0=V_0/\phi_0$, where the Newton's constant is given by $G_N=1/8\pi\phi_0$. One may incorrectly think that, in this latter limit $\phi=\phi_0$, an inconsistency arises due to equations \eqref{5} and \eqref{kg-bd-eq'}, according to which: $R=0$, and $T^{(m)}=4V_0$. Notice, however, that in the constant field limit, $\phi$ is not a dynamical variable and the equations \eqref{5}, \eqref{kg-bd-eq'} do not arise. In the limit $\omega_\text{BD}\rightarrow\infty$, $\alpha^2\rightarrow 0$, since the scalar field decouples from gravitation, the resulting theory is also general relativity.

%%%%%%%%%%%%%%%%%%%%%%%%%%%%%%%%%%%%%%%%%%%%%%%%%%%%%%%%%%%%%%
%%%%%%%%%%%%%%%%%%%%%%%%%%%%%%%%%%%%%%%%%%%%%%%%%%%%%%%%%%%%%%

\section{Scalar-tensor theories in general}\label{sec-stt-gen}

As an additional illustration of the generality of the approach exposed in this paper, which is the standard variational procedure used to derive the field equations in very well-known texts \cite{feynman, book, landau, gr-book, carroll, fujii}, let us choose a general scalar-tensor (ST) gravitational action of the form

\bea S_\text{ST}=\int_{\cal M} d^4x\sqrt{|g|}\left[f(\phi)R-K(\phi)(\der\phi)^2-V(\phi)\right],\label{st-action}\eea where $f(\phi)$ and $K(\phi)$ are functions of the $\phi$-field. Following the exposed procedures, the variation of this action with respect to the metric ($\phi$ fixed):

\bea \delta_g S_\text{ST}=\int_{\cal M} d^4x\sqrt{|g|}\left\{f(\phi)\delta R-K(\phi)\nabla_\mu\phi\nabla_\nu\phi\delta g^{\mu\nu}+\frac{\delta\sqrt{|g|}}{\sqrt{|g|}}\left[f(\phi)R-K(\phi)(\der\phi)^2-V(\phi)\right]\right\},\nonumber\eea can be written in the form (recall that $\delta R=\delta g^{\mu\nu}R_{\mu\nu}+g^{\mu\nu}\delta R_{\mu\nu}$):

\bea \delta_g S_\text{ST}=\int_{\cal M} d^4x\sqrt{|g|}\delta g^{\mu\nu}\left\{f(\phi)R_{\mu\nu}-K(\phi)\nabla_\mu\phi\nabla_\nu\phi-\frac{1}{2}g_{\mu\nu}\left[f(\phi)R-K(\phi)(\der\phi)^2-V(\phi)\right]\right\}+\delta_g S^*,\nonumber\eea where the piece of the variation $\delta_g S^*=\int d^4x\sqrt{|g|}f(\phi)g^{\mu\nu}\delta R_{\mu\nu}$, is to be compared either with the variation \eqref{var-bar-action} or with \eqref{fdr-2}. We get: $$\delta_g S^*=\int_{\cal M} d^4x\sqrt{|g|}f(\phi)g^{\mu\nu}\delta R_{\mu\nu}=\int_{\cal M} d^4x\sqrt{|g|}\delta g^{\mu\nu}\left(-\nabla_\mu\nabla_\nu+g_{\mu\nu}\Box\right)f(\phi).$$ The resulting Einstein-ST field equations are straightforwardly derived by requiring $\delta_g S_\text{ST}=0$: 

\bea f(\phi) G_{\mu\nu}=K(\phi)\left[\nabla_\mu\phi\nabla_\nu\phi-\frac{1}{2}g_{\mu\nu}(\der\phi)^2\right]-\frac{1}{2}g_{\mu\nu}V(\phi)+\left(\nabla_\mu\nabla_\nu-g_{\mu\nu}\Box\right)f(\phi)+\frac{T^{(m)}_{\mu\nu}}{2},\label{e-st-eq}\eea where, in addition to the action $S_\text{ST}$ in \eqref{st-action}, we have considered also a matter action $S_m$, whose variation: $\delta_g S_m=-\int d^4x\sqrt{|g|}\delta g^{\mu\nu}T^{(m)}_{\mu\nu}/2$, contributes a one half of the stress-energy tensor of matter in the RHS of the Einstein's equations. 

If vary the action \eqref{st-action} with respect to the scalar field (the $g_{\mu\nu}$ taken fixed): $$\delta_\phi S_\text{ST}=\int_{\cal M} d^4x\sqrt{|g|}\left[f'R\delta\phi-K'(\der\phi)^2\delta\phi-V'\delta\phi-2K\nabla^\mu\phi\nabla_\mu(\delta\phi)\right],$$ where the comma accounts for derivative with respect to the $\phi$-field, and we have to realize that: $$\nabla_\mu\left[K\nabla^\mu\phi\delta\phi\right]=K'(\der\phi)^2\delta\phi+K\Box\phi\delta\phi+\underline{K\nabla^\mu\phi\nabla_\mu(\delta\phi)},$$ then, after taking into account the Gauss theorem, one obtains the following Klein-Gordon-type equation:

\bea 2K\Box\phi+K'(\der\phi)^2+f'R=V',\label{st-kg-eq}\eea or, if additionally consider the trace of \eqref{e-st-eq}: $fR=K(\der\phi)^2+2V+3\Box f-T^{(m)}/2$ (notice that $\Box f=f''(\der\phi)^2+f'\Box\phi$), and substitute back into \eqref{st-kg-eq}, then

\bea \left[2K+3\frac{f'^2}{f}\right]\Box\phi+\left[K'+\frac{f'}{f}\left(K+3f''\right)\right](\der\phi)^2=V'-2\frac{f'}{f}V+\frac{f'}{2f}T^{(m)}.\label{st-kg-eq'}\eea The motion equations of the general scalar-tensor theory depicted by the action \eqref{st-action} are the equations \eqref{e-st-eq} and \eqref{st-kg-eq'}.

%%%%%%%%%%%%%%%%%%%%%%%%%%%%%%%%%%%%%%%%%%%%%%%%%%%%%%%

\subsection{Cubic self-interaction of the general form}

We want to make some remarks on the inclusion of a cubic self-interaction of the general form:

\bea S^\text{cubic}_{h(\phi)}=\int_{\cal M}d^4x\sqrt{|g|}\,h(\phi)\Box\phi(\der\phi)^2,\label{conclu-1}\eea where $h=h(\phi)$ is an arbitrary function of the $\phi$-field. Following the procedure below equation \eqref{var-3bic}, including equations \eqref{bound} and \eqref{var-g-cubic}, the variation of this action with respect to the metric,

\bea \delta_g S^\text{cubic}_{h(\phi)}=\int_{\cal M}d^4x\sqrt{|g|}\delta g^{\mu\nu}\left[-\frac{1}{2}g_{\mu\nu}h\Box\phi(\der\phi)^2+h\Box\phi\nabla_\mu\phi\nabla_\nu\phi+h(\der\phi)^2\nabla_\mu\nabla_\nu\phi\right]+\delta_g \bar{S}^\text{cubic}_{h(\phi)},\nonumber\eea where 

\bea \delta_g \bar{S}^\text{cubic}_{h(\phi)}=\int_{\cal M}d^4x\sqrt{|g|}\,h(\der\phi)^2\left[\nabla_\nu\phi\nabla_\mu(\delta g^{\mu\nu})-\frac{1}{2}g_{\mu\nu}\nabla_\lambda\phi\nabla^\lambda(\delta g^{\mu\nu})\right],\nonumber\eea can be written in the following way: $$\delta_g S^\text{cubic}_{h(\phi)}=-\frac{1}{2}\int_{\cal M}d^4x\sqrt{|g|}\delta g^{\mu\nu}T^{[h(\phi)]}_{\mu\nu},$$ where (check this as a training exercise)

\bea T^{[h(\phi)]}_{\mu\nu}=-2\left\{h\nabla_\mu\phi\nabla_\nu\phi\Box\phi-\nabla_{(\mu}\phi\nabla_{\nu)}\left[h(\der\phi)^2\right]+\frac{1}{2}g_{\mu\nu}\nabla_\lambda\phi\nabla^\lambda\left[h(\der\phi)^2\right]\right\},\label{conclu-2}\eea is the ``stress-energy'' tensor of the cubic self-interaction. We recall that, in order to meet the requirement of symmetry of $T^{[h(\phi)]}_{\mu\nu}$ under the exchange of the indexes $\mu\Leftrightarrow\nu$, the following equality have been taken into account: $$\delta g^{\mu\nu}\nabla_\nu\phi\nabla_\mu[h(\der\phi)^2]=\frac{1}{2}\delta g^{\mu\nu}\left\{\nabla_\nu\phi\nabla_\mu[h(\der\phi)^2]+\nabla_\mu\phi\nabla_\nu[h(\der\phi)^2]\right\}\equiv\delta g^{\mu\nu}\nabla_{(\mu}\phi\nabla_{\nu)}\left[h(\der\phi)^2\right].$$

Let us now to derive the contribution of the cubic term of the general form \eqref{conclu-1} towards the Klein-Gordon equation. In order to do so we have to vary \eqref{conclu-1} with respect to the $\phi$-field (the metric being held fixed):\footnote{Here we shall write the details of the derivation even if several of the steps are a repetition of the ones undertaken in Sect. \ref{subsec-3-1}.} 

\bea \delta_\phi S^\text{cubic}_{h(\phi)}=\int_{\cal M}d^4x\sqrt{|g|}\left[h'\Box\phi(\der\phi)^2\delta\phi+2h\Box\phi\nabla^\mu\phi\nabla_\mu(\delta\phi)+h\Box(\delta\phi)(\der\phi)^2\right],\label{cubic-x-1}\eea where, as before, the comma denotes derivative with respect to the $\phi$-field. If realize that, under the integral the divergence: $$\nabla^\mu\left[h(\der\phi)^2\nabla_\mu(\delta\phi)\right]=h'\nabla^\mu\phi\nabla_\mu(\delta\phi)(\der\phi)^2+h\nabla_\mu(\delta\phi)\nabla^\mu(\der\phi)^2+\underline{h\Box(\delta\phi)(\der\phi)^2},$$ vanishes, and substitute the underlined term into \eqref{cubic-x-1}, then we obtain

\bea \delta_\phi S^\text{cubic}_{h(\phi)}=\int_{\cal M}d^4x\sqrt{|g|}\left[h'\Box\phi(\der\phi)^2\delta\phi+W^\mu\nabla_\mu(\delta\phi)\right],\label{cubic-x-2}\eea where we have defined the vector $W^\mu\equiv 2h\Box\phi\nabla^\mu\phi-h'(\der\phi)^2\nabla^\mu\phi-h\nabla^\mu(\der\phi)^2$. Then, since $$\nabla_\mu(W^\mu\delta\phi)=\nabla_\mu W^\mu\delta\phi+\underline{W^\mu\nabla_\mu(\delta\phi)},$$ after integrating by parts we obtain the following expression for the variation of the action $S^\text{cubic}_{h(\phi)}$:

\bea \delta_\phi S^\text{cubic}_{h(\phi)}=\int_{\cal M}d^4x\sqrt{|g|}\delta\phi\left[h'\Box\phi(\der\phi)^2\delta\phi-\nabla_\mu W^\mu\right],\label{cubic-x-2'}\eea where

\bea \nabla_\mu W^\mu=h'\Box\phi(\der\phi)^2+2h(\Box\phi)^2-h''(\der\phi)^4-2h'\nabla^\mu\phi\nabla_\mu(\der\phi)^2+\underline{2h\nabla^\mu\phi\nabla_\mu(\Box\phi)-h\Box(\der\phi)^2},\label{conclu-div-1}\eea and the underlined terms contain the derivatives of higher (3rd) order. We have:

\bea &&\underline{2h\nabla^\mu\phi\nabla_\mu(\Box\phi)-h\Box(\der\phi)^2}=2h\left\{\left[\nabla_\mu(\Box\phi)-\Box(\nabla_\mu\phi)\right]\nabla^\mu\phi-\nabla^\mu\nabla^\nu\phi\nabla_\mu\nabla_\nu\phi\right\}\nonumber\\
&&\;\;\;\;\;\;\;\;\;\;\;\;\;\;\;\;\;\;\;\;\;\;\;\;\;\;\;\;\;\;\;\;\;\;\;\;\;\;\;\;\;\;\;=-2h\left[R_{\mu\nu}\nabla^\mu\phi\nabla^\nu\phi+\nabla^\mu\nabla^\nu\phi\nabla_\mu\nabla_\nu\phi\right],\label{cubic-x-3}\eea where in the last equality, thanks to the equation \eqref{nc-useful}, we have replaced the higher order derivatives by the term with the Ricci scalar: $R_{\mu\nu}\nabla^\mu\phi\nabla^\nu\phi$. If substitute the underlined terms in \eqref{cubic-x-3} into \eqref{conclu-div-1}, and then into \eqref{cubic-x-2'}, we can write the variation $\delta_\phi S^\text{cubic}_{h(\phi)}$ into the following form:

\bea \delta_\phi S^\text{cubic}_{h(\phi)}=\int_{\cal M}d^4x\sqrt{|g|}\delta\phi\left\{-2h(\Box\phi)^2+h''(\der\phi)^4+2h\nabla^\mu\nabla^\nu\phi\nabla_\mu\nabla_\nu\phi+2h\nabla^\mu\phi\nabla^\nu\phi\left[R_{\mu\nu}+2\frac{h'}{h}\nabla_\mu\nabla_\nu\phi\right]\right\}.\label{var-cub-fin}\eea It can be checked that, if set $h(\phi)=\alpha^2\phi^{-2}$ in \eqref{var-cub-fin}, one obtains \eqref{var-phi-cubic}.

%%%%%%%%%%%%%%%%%%%%%%%%%%%%%%%%%%%%%%%%%%%%%%%%%%%%%%
%%%%%%%%%%%%%%%%%%%%%%%%%%%%%%%%%%%%%%%%%%%%%%%%%%%%%%

\section{Several tips and an exercise}\label{sec-tips} 

In this section we leave as an exercise to obtain the Einstein's equations, as well as the Klein-Gordon equation, for the coupled Galileon theory which is given by the following action \cite{ex}: 

\bea S_\text{cG}=\int d^4x\sqrt{|g|}\left[\left(1-2k\frac{\phi}{M_\text{pl}}\right)\frac{M^2_\text{pl}}{2}R-\frac{1}{2}(\der\phi)^2-\frac{\alpha}{2M^3_\text{pl}}(\der\phi)^2\Box\phi-V(\phi)+{\cal L}_m\right],\label{cg-action}\eea where $M^2_\text{pl}=(8\pi G_N)^{-1}$, and $\alpha$ is a free constant. As before ${\cal L}_m$ stands for the Lagrangian of the matter degrees of freedom other than the Galileon itself. Notice that the factor of $2$ in the matter term has been omitted. We give several tips:

\begin{itemize}

\item First, split the action into a scalar-tensor part: 

\bea S_\text{ST}=\int d^4x\sqrt{|g|}\left[f(\phi)R-K(\phi)(\der\phi)^2-V(\phi)\right],\label{st-cg-action}\eea where 

\bea f(\phi)=\left(1-2k\frac{\phi}{M_\text{pl}}\right)\frac{M^2_\text{pl}}{2},\;K(\phi)=\frac{1}{2},\label{replacement}\eea a piece with the cubic self-interaction 

\bea S_\text{int}=-\frac{\alpha}{M^3}\int d^4x\sqrt{|g|}(\der\phi)^2\Box\phi,\label{int-act}\eea and a matter piece $S_m=\int d^4x\sqrt{|g|}{\cal L}_m$.

\item Variation of the action $S_\text{ST}+S_m,$ with respect to the metric $g_{\mu\nu}$, leads to the equations \eqref{e-st-eq} with the appropriate replacements given in \eqref{replacement}, where we have to realize that in the present case: $$\left(\nabla_\mu\nabla_\nu-g_{\mu\nu}\Box\right)f(\phi)=-kM_\text{pl}\left(\nabla_\mu\nabla_\nu-g_{\mu\nu}\Box\right)\phi.$$ We obtain:

\bea \left(1-2k\frac{\phi}{M_\text{pl}}\right)G_{\mu\nu}=\frac{1}{M^2_\text{pl}}\left[\nabla_\mu\phi\nabla_\nu\phi-\frac{1}{2}g_{\mu\nu}(\der\phi)^2-g_{\mu\nu}V(\phi)\right]-\frac{2k}{M_\text{pl}}\left(\nabla_\mu\nabla_\nu-g_{\mu\nu}\Box\right)\phi+\frac{T^{(m)}_{\mu\nu}}{M^2_\text{pl}}.\label{e-st-eq-ex}\eea 

\item Variation of the above action $S_\text{ST}+S_m$, with respect to the $\phi$-field keeping the metric fixed, yields the STKG equation \eqref{st-kg-eq}, with the above mentioned replacements for $f(\phi)$ and $K(\phi)$ in \eqref{replacement}:

\bea \Box\phi-kM_\text{pl}R=V',\label{st-kg-eq-ex}\eea or, if substitute the curvature scalar $R$ from the trace of \eqref{e-st-eq-ex} into \eqref{st-kg-eq-ex}, then we obtain the STKG equation in the form of \eqref{st-kg-eq'} with the replacements of \eqref{replacement}:

\bea \left(1+6k^2-2k\frac{\phi}{M_\text{pl}}\right)\Box\phi-\frac{k}{M_\text{pl}}(\der\phi)^2=\left(1-2k\frac{\phi}{M_\text{pl}}\right)V'+4\frac{k}{M_\text{pl}}V-\frac{k}{M_\text{pl}}T^{(m)},\label{st-kg-eq-ex'}\eea

\item Variation of the action piece $S_\text{int}$ in \eqref{int-act}, with respect to the metric yields \eqref{conclu-2} with the substitution $h(\phi)=-\alpha/2M^3_\text{pl}$:

\bea T^{(\alpha)}_{\mu\nu}=\frac{\alpha}{M^3_\text{pl}}\left\{\nabla_\mu\phi\nabla_\nu\phi\Box\phi-\nabla_{(\mu}\phi\nabla_{\nu)}(\der\phi)^2+\frac{1}{2}g_{\mu\nu}\nabla_\lambda\phi\nabla^\lambda(\der\phi)^2\right\}.\label{set-int-ex}\eea This stress-energy tensor of the cubic self-interaction (times $M_\text{pl}^{-2}$) is to be added in the RHS of \eqref{e-st-eq-ex}.

Variation of $S_\text{int}$ with respect to the $\phi$-field leads to \eqref{var-cub-fin} with $h(\phi)=-\alpha/2M^3_\text{pl}$. The terms in curly brackets under the integral in \eqref{var-cub-fin} with the mentioned substitution:

\bea \frac{\alpha}{M^3_\text{pl}}\left[(\Box\phi)^2-\nabla^\mu\nabla^\nu\phi\nabla_\mu\nabla_\nu\phi-\nabla^\mu\phi\nabla^\nu\phi R_{\mu\nu}\right],\label{curly-ex}\eea are to be added in the LHS of \eqref{st-kg-eq-ex} (or of \eqref{st-kg-eq-ex'}).

\end{itemize}

By straightforwardly applying the procedure explained in this paper, as well as by following the above tips, we are led to the answer to the proposed exercise:

\bea &&\left(1-2k\frac{\phi}{M_\text{pl}}\right)G_{\mu\nu}=\frac{1}{M^2_\text{pl}}\left[T^{(\phi)}_{\mu\nu}+T^{(\alpha)}_{\mu\nu}+T^{(m)}_{\mu\nu}\right]-\frac{2k}{M_\text{pl}}\left(\nabla_\mu\nabla_\nu-g_{\mu\nu}\Box\right)\phi,\nonumber\\
&&\Box\phi+\frac{\alpha}{M^3_\text{pl}}\left[(\Box\phi)^2-\nabla_\mu\nabla_\nu\phi\nabla^\mu\nabla^\nu\phi-\nabla^\mu\phi\nabla^\nu\phi R_{\mu\nu}\right]-kM_\text{pl}R=V',\label{conclu-3}\eea where $$T^{(\phi)}_{\mu\nu}=\nabla_\mu\phi\nabla_\nu\phi-\frac{1}{2}g_{\mu\nu}(\der\phi)^2-g_{\mu\nu}V(\phi),$$ is the stress-energy tensor of the Galileon, and the stress-energy tensor of the cubic self-interaction: $T^{(\alpha)}_{\mu\nu}$, is given by Eq. \eqref{set-int-ex}.

%%%%%%%%%%%%%%%%%%%%%%%%%%%%%%%%%%%%%%%%%%%%%%%%%%%%%%%%%%%%%%%%%%%%%%%%%%%%%%%%%%%%%%%%
%%%%%%%%%%%%%%%%%%%%%%%%%%%%%%%%%%%%%%%%%%%%%%%%%%%%%%%%%%%%%%%%%%%%%%%%%%%%%%%%%%%%%%%%

\section{The BD-Galileon cosmological model with the cubic interaction}\label{sec-cosmo} 

In order to put the above explained procedure and the obtained equations of motion in a context of interest for the applications, in this section we shall write the equations for an specific spacetime model: the Friedmann-Robertson-Walker spacetime, and for a specific cosmological model: the BD-Galileon cosmological model with the cubic interaction, which is based in the equations of motion \eqref{e-bd-eq'} and \eqref{kg-bd-eq'}, that are derived from the action \eqref{kazuya-action}. Several very well-known expressions of frequent use will be given so that non-cosmologists can make their own calculations. For completeness of the exposition, a brief discussion on the main physical features of the chosen cosmological model will be also provided.

%%%%%%%%%%%%%%%%%%%%%%%%%%%%%%%%%%%%%%%%%%%%%

\subsection{The cosmological field equations}

Here we assume the FRW spacetime with flat spatial sections, whose line element is given by 

\bea ds^2=-dt^2+a^2(t)\delta_{ik}dx^idx^k,\;\;\;i,k=1,2,3,\label{frw-line}\eea where $a=a(t)$ is the scale factor: a dimensionless function of the cosmic time $t$, and the three-dimensional Kronecker delta $\delta_{ik}=1$ if $i=k$ (it is vanishing otherwise). For sake of simplicity we imagine a cosmological background filled with a perfect fluid with stress-energy tensor: 

\bea T^{(m)}_{\mu\nu}=\left(\rho_m+p_m\right)u_\mu u_\nu+p_m g_{\mu\nu},\label{pf-set}\eea where $\rho_m$ is the energy density of the background matter, $p_m$ is its barotropic pressure with equation of state $p_m=w_m\rho_m$, and $u_\mu=\delta^0_\mu$ is the 4-velocity of co-moving observers (free falling observers, i. e., observers moving with the expansion, so that the only non-vanishing component of the 4-velocity is $u_0=\delta^0_0=1$). The resulting Einstein-BD field equations \eqref{e-bd-eq'} read ($H\equiv\dot a/a$ is the Hubble parameter):

\bea &&3H^2=-3HP+\frac{\omega_\text{BD}}{2}P^2+\frac{V}{\phi}+\alpha^2\left(3HP^3+P^4\right)+\frac{\rho_m}{\phi},\label{fried-eq}\\
&&2\dot H+3H^2=-2HP-\dot P-P^2-\frac{\omega_\text{BD}}{2}P^2+\frac{V}{\phi}+\alpha^2P^2\dot P-\frac{p_m}{\phi},\label{raucha-eq}\eea where, following \cite{kazuya}, we have introduced the new dynamical variable $P\equiv\dot\phi/\phi$, while the KGBD equation \eqref{kg-bd-eq'} amounts to:

\bea &&Q_6\left(\dot P+P^2\right)=-3Q_0HP-2\gamma_\text{BD}\alpha^2\left(3\dot HP^2-2P^4\right)-2\gamma_\text{BD}\frac{\phi\der_\phi V-2V}{\phi}+\gamma_\text{BD}\frac{\rho_m-3p_m}{\phi},\label{kg-eq}\eea where $\gamma_\text{BD}=(3+2\omega_\text{BD})^{-1}$, and we have defined the parametric family of functions

\bea Q_l\equiv Q_l(H,P):=1+\gamma_\text{BD}\alpha^2\left[(l+6)HP+(l-1)P^2\right].\label{qdefi}\eea 

An appropriate combination of the equations \eqref{fried-eq} and \eqref{raucha-eq} -- as a matter of fact it is their sum -- leads to:

\bea 2\dot H+6H^2=-5HP-\left(1-\alpha^2P^2\right)\left(\dot P+P^2\right)+3\alpha^2HP^3+2\frac{V}{\phi}+\frac{\rho_m-p_m}{\phi}.\label{combi-eq}\eea Written in this form the equation \eqref{combi-eq} can be useful, for instance, for the dynamical systems study of the cosmological model based on \eqref{kazuya-action}.

%%%%%%%%%%%%%%%%%%%%%%%%%%%%%%%%

\subsection{Useful expressions}

In the above equations we have considered that, consistently with the FRW cosmological spacetime model, the BD scalar field is a function of the cosmic time $t$ only: $\phi=\phi(t)$. In this sense the gravitational coupling $G\propto\phi^{-1}$, is also an evolving function of the cosmic time. Besides, in order to write the motion equations \eqref{e-bd-eq'} and \eqref{kg-bd-eq'} in terms of the FRW metric above, we have taken into account the following very well-known, frequently used expressions which are valid only for the FRW metric with flat spatial sections (the $\Gamma^\alpha_{\;\mu\nu}$ are the Christoffel symbols of the metric):

\bea &&\Gamma^\alpha_{\;\mu\nu}\neq 0\;\;\Rightarrow\;\;\Gamma^0_{\;ij}=a^2H\delta_{ij},\;\Gamma^i_{\;0j}=\Gamma^i_{\;j0}=H\delta^i_j,\nonumber\\
&&(\der\phi)^2=-\dot\phi^2=-\phi^2P^2,\;\Box\phi=-\ddot\phi-3H\dot\phi=-\phi\left(\dot P+P^2+3HP\right),\nonumber\\
&&\nabla_0\nabla_0\phi=\ddot\phi=\phi\left(\dot P+P^2\right),\;\nabla_i\nabla_j\phi=-a^2H\dot\phi\delta_{ij}=-a^2\phi HP\delta_{ij},\nonumber\\
&&R_{00}=-3\dot H-3H^2,\;R_{ij}=a^2\left(\dot H+3H^2\right)\delta_{ij},\;R=6\left(\dot H+2H^2\right),\nonumber\\
&&G_{\mu\nu}=R_{\mu\nu}-\frac{1}{2}g_{\mu\nu}R\neq 0\;\;\Rightarrow\;\;G_{00}=3H^2,\;G_{ij}=-a^2\left(2\dot H+3H^2\right)\delta_{ij},\label{freq-use}\eea while for the non-null components of the stress-energy tensors \eqref{def-1}, \eqref{cub-sec} and \eqref{pf-set}, respectively, we get:

\bea &&T^{(\phi)}_{00}=\phi\left[\frac{\omega_\text{BD}}{2}P^2+\frac{V(\phi)}{\phi}\right],\;T^{(\phi)}_{ij}=\phi a^2\left[\frac{\omega_\text{BD}}{2}P^2-\frac{V(\phi)}{\phi}\right]\delta_{ij},\nonumber\\
&&T^{(\alpha^2)}_{00}=\phi\alpha^2\left(3HP^3+P^4\right),\;T^{(\alpha^2)}_{ij}=-\phi a^2\alpha^2P^2\dot P\delta_{ij},\nonumber\\
&&T^{(m)}_{00}=\rho_m,\;T^{(m)}_{ij}=a^2p_m\delta_{ij}.\label{freq-use-1}\eea 

We leave as a training exercise for non-cosmologists to write the motion equations \eqref{conclu-3} in terms of the FRW metric with flat spatial sections \eqref{frw-line}.

%%%%%%%%%%%%%%%%%%%%%%%%%%%%%%%%%%%%%%%%%%%%%%%%%%%%%%%%%%%%%%%%%%%%%%%%%%%%%%

\subsection{Comments on the BD-Galileon model with the cubic self-interaction}

We want to make a couple of comments on the present cosmological model: the Brans-Dicke Galileon \cite{kazuya}. The model shares certain similitude with the 5-dimensional Dvali-Gabadadze-Porrati (DGP) model \cite{dgp, cosmo-dgp}.\footnote{This model describes a brane with 4D world-volume, that is embedded into a flat 5D bulk, and allows for infrared -- large scale -- modifications of the gravitational laws. A distinctive ingredient of the model is the induced Einstein-Hilbert action on the brane, that is responsible for the recovery of 4D Einstein gravity at moderate scales, even if the mechanism of this recovery is rather non-trivial. The acceleration of the expansion at late times is explained here as a consequence of the leakage of gravity into the bulk at large (cosmological) scales, so, it is just a 5D geometrical effect unrelated to any kind of mysterious ``dark energy''.} Actually:

\begin{itemize}

\item The Friedmann equation \eqref{fried-eq}, that can be written in the following way: 

\bea H^2=\frac{\rho_m}{3\phi}-HP(1-\alpha^2P^2)+\frac{\omega_\text{BD}}{6}P^2+\frac{V}{3\phi}+\frac{\alpha^2}{3}P^4,\label{dgp-fried-eq}\eea allows for two branches of
solutions, depending on the sign of $P=\dot\phi/\phi$. In analogy, the modified Friedmann equation in DGP \cite{cosmo-dgp} also has two branches: 

\bea H^2=\frac{\rho_m}{3M^2_\text{pl}}\pm\frac{1}{r_c}H,\label{dgp-eq}\eea where the constant parameter $r_c$ is the so called cross-over distance.

\item The study of the stability of the present model against small perturbations $\phi\rightarrow\phi+\delta\phi$, yields that the perturbations are stable for positive $P$, while, for negative $P<0$, ghostlike instabilities arise. In a similar fashion the '+' branch of the DGP cosmological model is unstable, while the '-' branch is stable instead. 

\item In a similar way than in the DGP model,\footnote{In DGP, for the '+' branch of \eqref{dgp-eq}, as long as the expansion proceeds and the energy density of matter consequently dilutes ($\rho_m\propto a^{-3}$), at late times one gets a de Sitter solution $H=1/r_c$, called as ``self-accelerating'' because one does not need of any kind of exotic (or non-exotic) matter in order to have accelerated expansion.} a self-accelerating solution can be found in the BD-Galileon model with the cubic self-interaction \cite{kazuya}. This solution can be obtained if in the motion equations \eqref{fried-eq}, \eqref{raucha-eq} and \eqref{kg-eq}, set $\dot H=\dot P=0$, and $V=\rho_m=p_m=0$. Unlike the self-accelerating solution in the '+' branch of DGP which is generic, the self-accelerating solution in the BD Galileon with the cubic self-interaction exists if $\omega_\text{BD}<-4/3$. As a matter of fact, under the above requirements, the equation \eqref{raucha-eq} can be written as $$H^2+\frac{2}{3}PH+\frac{P^2}{3}\left(1+\frac{\omega_\text{BD}}{2}\right)=0\;\Rightarrow\;H_\pm=\frac{P}{3}\left(-1\pm\sqrt{-\frac{3\omega_\text{BD}}{2}-2}\right).$$ Since we have to choose the '+' branch solution $H_+$ in order for the universe to be expanding, then, as a matter of fact a more stringent bound $\omega_\text{BD}<-2$ is to be met by the self-accelerating solution. Additionally, in order to meet the requirement that the present stage of the cosmic expansion be accelerating, it is required that $\alpha\sim H_0^{-1}$.

\end{itemize}

In order to recover GR at early times in this model, $P$ should be smaller than $H$. In fact, GR is recovered in the limit $P\rightarrow 0$. This is indeed achieved at high energies when the nonlinear terms in the field equation dominate over the linear term \cite{kazuya}. This is a cosmological version of the Vainshtein mechanism to screen the scalar field \cite{chow}.

%%%%%%%%%%%%%%%%%%%%%%%%%%%%%%%%%%%%%%%%%%%%%%%%%%%%%
%%%%%%%%%%%%%%%%%%%%%%%%%%%%%%%%%%%%%%%%%%%%%%%%%%%%%

\section{Conclusion}\label{sec-conclu}

In this pedagogically oriented paper we have performed a detailed and systematic explanation of the way the equations of motion can be derived by means of the variational principle from the action of the Brans-Dicke theory with the cubic self-interaction \eqref{kazuya-action}. This theory is one of the simplest modifications of general relativity, yet, it preserves several of the complications which are typical of more complex modifications such as the $f(R)$-theories. Besides, the procedures and results of this work are straightforwardly applicable to the latter theories (see Sec. \ref{subsec-2-2}). The case with the scalar-tensor theories in general has been also developed in detail in Sec. \ref{sec-stt-gen}. The derivation of the motion equations from an action taken from the bibliography: the one corresponding to the so called coupled Galileon \cite{ex}, has been left as an exercise to the readers. Several tips have been provided to help the readers straightforwardly get to the correct answer.

Whenever it has been possible, we have been able to show how the higher order derivatives of the $\phi$-field, appearing in the intermediate computations, are harmless since these are canceled out by similar terms with the opposite sign in the final steps. As a matter of fact, what we do is to trade the terms with the 3rd order derivatives in $\phi$ by a term containing the Ricci tensor \eqref{nc-useful}: $\Box(\nabla_\mu\phi)-\nabla_\mu(\Box\phi)=R_{\mu\nu}\nabla^\nu\phi.$ This results in that, as expected \cite{nicolis, deffayet, horndeski}, the final equations of motion are 2nd order in the derivatives. 

Here we have underlined, also, several subtleties associated with the variation of a piece of action containing higher order derivatives of the field, which may go unnoticed even to experts. For instance, while taking the variation of a term containing $\Box\phi(\der\phi/\phi)^2$ with respect to the metric: $$\delta_g\Box\phi\left(\frac{\der\phi}{\phi}\right)^2=\delta g^{\mu\nu}\frac{\nabla_\mu\phi}{\phi}\frac{\nabla_\nu\phi}{\phi}\Box\phi+\left(\frac{\der\phi}{\phi}\right)^2\delta\Box\phi,$$ in the variation of the second term in the RHS of the above equation, the correct variation is: $\delta\Box\phi=\delta g^{\mu\nu}\nabla_\mu\nabla_\nu\phi-g^{\mu\nu}\nabla_\lambda\phi\delta\Gamma^\lambda_{\;\mu\nu}$, and not $\delta\Box\phi=\delta g^{\mu\nu}\nabla_\mu\nabla_\nu\phi$, i. e., one has to take into account the correct definition of the covariant derivative of the vector field $\der_\mu\phi$, so that the Christoffel symbols also contribute towards the variation of $\Box\phi$ with respect to he metric. This is a very known, almost trivial fact, yet it may go unnoticed. 

Another subtlety to take into account is related with the variation of the cubic interaction term \eqref{conclu-1} with respect to the metric: $$\delta_g S^\text{cubic}_{h(\phi)}=-\frac{1}{2}\int_{\cal M}d^4x\sqrt{|g|}\delta g^{\mu\nu}T^{[h(\phi)]}_{\mu\nu},$$ where $T^{[h(\phi)]}_{\mu\nu}$, which is given by \eqref{conclu-2}, is a symmetric tensor since it appears in the RHS of the Einstein's equations. While taking the variation $\delta_g S^\text{cubic}_{h(\phi)}$, it appears the term $\delta g^{\mu\nu}\nabla_\nu\phi\nabla_\mu[h(\der\phi)^2]$. The subtlety has to do with the fact that, in order to meet the requirement of symmetry of $T^{[h(\phi)]}_{\mu\nu}$ under the exchange of the indexes $\mu\Leftrightarrow\nu$, it is mandatory to take into account the following equality: $$\delta g^{\mu\nu}\nabla_\nu\phi\nabla_\mu[h(\der\phi)^2]=\frac{1}{2}\delta g^{\mu\nu}\left\{\nabla_\nu\phi\nabla_\mu[h(\der\phi)^2]+\nabla_\mu\phi\nabla_\nu[h(\der\phi)^2]\right\}\equiv\delta g^{\mu\nu}\nabla_{(\mu}\phi\nabla_{\nu)}\left[h(\der\phi)^2\right].$$ Hence, instead of $\nabla_\nu\phi\nabla_\mu[h(\der\phi)^2]$, it is $\nabla_{(\mu}\phi\nabla_{\nu)}\left[h(\der\phi)^2\right]$, the term that appears in the expression of $T^{[h(\phi)]}_{\mu\nu}$ in \eqref{conclu-2}.

In Sec. \ref{sec-cosmo}, in order to put the procedure and the results exposed in the present paper in a context of interest for applications, we have written the motion equations \eqref{e-bd-eq'}, \eqref{kg-bd-eq'} of the EBD-theory depicted by the action \eqref{kazuya-action}, in terms of the cosmological FRW metric with flat spatial sections. This is what is known as the BD Galileon model \cite{kazuya}. Several very well-known expressions of frequent use in the cosmological context were given -- see \eqref{freq-use} and \eqref{freq-use-1} -- so that the reader could find in one place frequently used unalike information. For completeness of the exposition, a brief discussion on the main physical features of the BD-Galileon cosmological model with the cubic interaction has been also provided.

We hope that graduate and postgraduate students can take advantage of this paper and that, after reading it, the derivation of the field equations starting from an action principle can be a useful and instructive exercise to them. The benefits from learning how the variational principle works for modifications of general relativity would enlighten the way towards an independent research path.

%%%%%%%%%%%%%%%%%%%%%%%%%
%%%%%%%%%%%%%%%%%%%%%%%%%

\section{acknowledgment}

The authors acknowledge very useful comments by Bartolome Alles, Kazuya Koyama and Sergei Odintsov. They are also grateful to SNI-CONACyT for continuous support of their research activity. The work of R G-S was partially supported by SIP20150188, SIP20160512, COFAA-IPN, and EDI-IPN grants. I Q and T G thank CONACyT of M\'exico for support of this research.

%%%%%%%%%%%%%%%%%%%%%%%%%%%%

\end{document}